\pdfoutput=1
\RequirePackage{currfile}
\documentclass[cernpreprint, atlasdraft=false, texlive=2016, USenglish, texmf]{atlasdoc}
 
\usepackage{atlaspackage}
\usepackage[block=space]{atlasbiblatex}
 
\usepackage[BSM]{atlasphysics}
 
\graphicspath{{logos/}{figures/}}
 
\usepackage{ANA-SUSY-2019-18-PAPER-defs}
\newcommand{\inmath}[1]{\ensuremath{#1}\xspace}
\newcommand{\eg}{e.g.\@\xspace}               
               
\newcommand{\ie}{i.e.\@\xspace}                   
                  
\newcommand{\percent}[1]{{#1}\%}

\newcommand{\Wjets}{\inmath{W+\textnormal{\,jets}}}
\newcommand{\Zjets}{\inmath{Z\,+\textnormal{\,jets}}}
\newcommand{\Vjets}{\inmath{V\,+\textnormal{\,jets}}}
\newcommand{\ttV}{\inmath{\ttbar\!\,+V}}
\newcommand{\ttX}{\inmath{\ttbar\!\,+X}}
\newcommand{\ttH}{\inmath{\ttbar + H}}

\newcommand{\ttFake}{{\ttbar}-fake\xspace}
\newcommand{\ttOneReal}{{\ttbar} (1~real~\inmath{\tau})\xspace}
\newcommand{\ttTwoReal}{{\ttbar} (2~real~\inmath{\tau})\xspace}

\newcommand{\GeVx}[1]{\unit[#1]{GeV}\xspace}
\newcommand{\TeVx}[1]{\unit[#1]{TeV}\xspace}

\newcommand{\ifbx}[1]{\inmath{\unit[#1]{fb^{-1}}}} 
\newcommand{\intlumi}{\ifbx{139}}
 
\newcommand{\CLs}{\inmath{\text{CL}_\text{s}}}               
\newcommand{\CLb}{\inmath{\text{CL}_\text{b}}}               
 
\newcommand{\sigmavis}{\inmath{\langle A \epsilon{\sigma} \rangle_\text{obs}^{95}}}
 
\newcommand{\sntev}[1]{\inmath{\sqrt{s} = \unit[#1]{\TeV}}}
\newcommand{\eighttev}{\sntev{8}}
\newcommand{\tttev}{\sntev{13}}
\newcommand{\cme}{\tttev}
 
\renewcommand{\stop}{\inmath{\skew{3}{\tilde}{t}}}

\newcommand{\stopnoone}{\stop}
\newcommand{\staunoone}{\stau}
 
\newcommand{\LQ  }{\inmath{\text{L\!Q}}}
\newcommand{\LQu }{\inmath{\text{L\!Q}_{3}^{\text{u}}}}
\newcommand{\LQd }{\inmath{\text{L\!Q}_{3}^{\text{d}}}}
\newcommand{\LQud}{\inmath{\text{L\!Q}_{3}^{\text{u/d}}}}
\newcommand{\LQv }{\inmath{\text{L\!Q}_{3}^{\text{v}}}}

\newcommand{\gravitino}{\gravino}
 
\newcommand{\mtsumbb}{\inmath{\sum\mt(b_{1,2})}}
\newcommand{\st}{\inmath{s_{\text{T}}}}
\newcommand{\mT     }{\inmath{m_{\text{T}}}}

\newcommand{\mt     }{\mT}
\newcommand{\mTtwo  }{\inmath{m_{\text{T2}}}}
\newcommand{\mttwo  }{\mTtwo}

\newcommand{\vecmet }{\inmath{\vec{E}_{\text{T}}^{\text{miss}}}}
\newcommand{\mtautau}{\inmath{m_\text{vis}(\tau_1,\tau_2)}}
\newcommand{\dRy}{\inmath{\Delta R_y}}

\renewcommand{\vec}[1]{\mathbf{#1}} 
\newcommand{\vecpt}{\inmath{\vec{p}_\text{T}}}
\newcommand{\vecqt}{\inmath{\vec{q}_{\text{T}}}}
 
\newcommand*{\PYTHIAV}[1]{\PYTHIA~#1\xspace}
\newcommand*{\POWHEGBOXV}[1]{\POWHEGBOX~#1\xspace}
\newcommand*{\SHERPAV}[1]{\SHERPA~#1\xspace}
\newcommand*{\MGMCatNLOV}[1]{\MGNLO~#1\xspace}
\newcommand*{\MGMCatNLO}{\MGNLO}
\newcommand*{\HERWIGV}[1]{\HERWIG~#1\xspace}

\newcommand{\todo}[1]{{\color{red}\emph{#1}}}
\newcommand{\massof}[1]{\inmath{m(#1)}}
\newcommand{\btagging}{\btag}

\newcommand{\BRcharged }{\inmath{B(\LQud \to q\ell)}}

\newcommand{\BRchargedv}{\inmath{B(\LQv  \to b\tau)}}
 
\newcommand{\ourlimitsLQ}{\TeVx{1.25}}
\newcommand{\ourlimitvLQYM}{\TeVx{1.8}}
\newcommand{\ourlimitvLQMC}{\TeVx{1.5}}
 
\newcommand{\paperornote}{paper\xspace}
 
\newcommand{\StackedHisto}{The stacked histograms show the various SM background contributions.\xspace}
\newcommand{\BandIsStatSyst}{The hatched band indicates the total statistical and systematic uncertainty of the SM background.\xspace}
\newcommand{\Postfit}{The \ttTwoReal and \ttOneReal contributions and the single-top background contributions are scaled with the normalization factors obtained from the background-only fit.\xspace}
\newcommand{\PostfitSingletau}{The \ttOneReal and single-top background contributions are scaled with the normalization factors obtained from the background-only fit.\xspace}
\newcommand{\RightBinOverflow}{The rightmost bin includes the overflow.\xspace}
\newcommand{\RightBinOverflowC}[1]{The rightmost bin includes the overflow#1.\xspace}
\newcommand{\MCdatascaled}{The predictions for the top-quark backgrounds are scaled with the normalization factors obtained from the background-only fit.\xspace}
\newcommand{\MCdatagood}{From the plots it can be seen that the background model describes the data very well.\xspace}
\newcommand{\UncertaintiesDoNotSumUp}{Since the individual uncertainties are correlated, they do not sum in quadrature to equal the total background uncertainty.\xspace}
\newcommand{\BackgroundOtherInPlots}[1]{
Minor backgrounds are grouped together and denoted by ``Other''.
This includes #1.\xspace
}
\AtlasTitle{
Search for new phenomena 
in $p\!p$~collisions 
in \\
final states with tau leptons, $b$-jets,
and missing transverse momentum
with the ATLAS detector
}
 
\AtlasAbstract{
A search for new phenomena 
in final states with hadronically decaying tau leptons, $b$-jets, and missing transverse momentum is presented.
The analyzed dataset comprises $p\!p$~collision data at a center-of-mass energy of \cme with an integrated luminosity of \intlumi,
delivered by the Large Hadron Collider
and recorded with the ATLAS detector from 2015 to 2018.
The observed data are compatible with the expected Standard Model background.
The results are interpreted in simplified models for two different scenarios.
The first model is based on supersymmetry and considers pair production of top squarks,
each of which decays into a $b$-quark, a neutrino and a tau slepton.
Each tau slepton in turn decays into a tau lepton and a nearly massless gravitino.
Within this model,
top-squark masses up to \TeVx{1.4} can be excluded at the \percent{95} confidence level over a wide range of tau-slepton masses.
The second model considers pair production of leptoquarks
with decays into third-generation leptons and quarks.
Depending on the branching fraction into charged leptons,
leptoquarks with masses up to around \ourlimitsLQ can be excluded at the \percent{95} confidence level
for the case of scalar leptoquarks and
up to \ourlimitvLQYM (\ourlimitvLQMC) for vector leptoquarks 
in a Yang--Mills (minimal-coupling) scenario.
In addition, model-independent upper limits are set on the cross section of processes beyond the Standard Model.
}
 
\AtlasJournal{Phys. Rev. D}
\AtlasJournalRef{\PRD 104 (2021) 112005}
\AtlasDOI{10.1103/PhysRevD.104.112005}
 
\PreprintIdNumber{CERN-EP-2021-118}
\hypersetup{pdftitle={ATLAS document},pdfauthor={The ATLAS Collaboration}}
 
\begin{document}
 
\maketitle
 
\tableofcontents

\section{Introduction}
\label{sec:intro}
% The next lines are included from the .//tex/introduction.tex input file
 
The Standard Model (SM) of particle physics has been verified to high precision.
Despite its success, several observations have been made
which have exposed the theory's shortcomings in various aspects and fostered new theoretical ideas.
Supersymmetry (SUSY)~\cite{Golfand:1971iw,Volkov:1973ix,Wess:1974tw,Wess:1974jb,Ferrara:1974pu,Salam:1974ig,Martin:1997ns} is a framework for models
that extend the symmetries underlying the SM by introducing superpartners of the known bosons and fermions
with the same quantum numbers but a spin difference of half a unit.
These models can address 
the gauge hierarchy problem~\cite{Sakai:1981gr,Dimopoulos:1981yj,Ibanez:1981yh,Dimopoulos:1981zb}.
When conservation of $R$-parity~\cite{Farrar:1978xj} is assumed,
the lightest supersymmetric particle is stable
and may provide a candidate particle for the cold dark matter component of the universe~\cite{Goldberg:1983nd,Ellis:1983ew}.
The introduction of supersymmetric partner particles can also modify the renormalization group equations in such a way
that the coupling constants of the SM electromagnetic, weak and strong interactions meet at one point at some high energy scale
as expected in a grand unified theory~\cite{DEBOER1994201}. 
Another possible way to extend the SM is
to embed the SM symmetry group in an overarching symmetry group,
such as SU(5)~\cite{georgi_glashow_unification} in grand unification,
which gives rise to a new class of bosons
that carry nonzero baryon and lepton quantum numbers and are charged under all SM gauge groups.
These 
leptoquarks (\LQ), which
can be either scalar or vector bosons, 
appear in a variety of SM extensions~\cite{FARHI1981277,SCHREMPP1985101,string,comp,Barbier:2004ez}
and would provide an explanation for the structural similarities of the quark and lepton sectors in the SM. 
Processes mediated by the exchange of leptoquarks can violate lepton-flavor universality
and have been proposed as an explanation~\cite{Hiller:2014yaa,Gripaios:2014tna,Freytsis:2015qca,Bauer:2015knc,DiLuzio:2017chi,Buttazzo:2017ixm,Cline:2017aed} 
for the deviations from the SM predictions seen by many experiments
in measurements of $B$-meson decays~\cite{babar,belle,lhcb1,Hirose:2016wfn, Aaij:2017tyk, Aaij:2015oid,Wehle:2016yoi,lhcb2,Aaij:2021vac}.
Contributions arising from leptoquarks with additional couplings to the muon could also bridge the gap~\cite{PhysRevD.95.055018,Chakraverty_2001}
between the theoretical prediction for the anomalous magnetic moment of the muon $(g-2)_\mu$ within the SM
and the experimentally measured value, which is higher by $4.2\sigma$~\cite{PhysRevLett.126.141801}.
 
In this \paperornote, a search for physics beyond that described in the Standard Model is conducted
using events with final states with one or more hadronically decaying tau leptons, one or more \btagged jets and large missing transverse momentum.
This is a signature that is sensitive to models in which the new particles preferentially decay into third-generation SM particles.
Two benchmark signal models are studied.
The first model considers the production of supersymmetric partner states of the third-generation SM particles,
while the second model foresees scalar leptoquarks that decay into third-generation SM particles.
An additional interpretation, for which the analysis was not explicitly optimized, is provided for vector leptoquarks that decay into third-generation SM particles.
The full Run-2 dataset of proton--proton (\textit{pp}) collisions recorded with the ATLAS detector
at the Large Hadron Collider (LHC) is analyzed.
This dataset corresponds to an integrated luminosity of $\SI{139}{\ifb}$, taken from 2015 through 2018,
at a center-of-mass energy of \cme.
 
\begin{figure}
\centering
\hfill
\includegraphics[width=4.2cm]{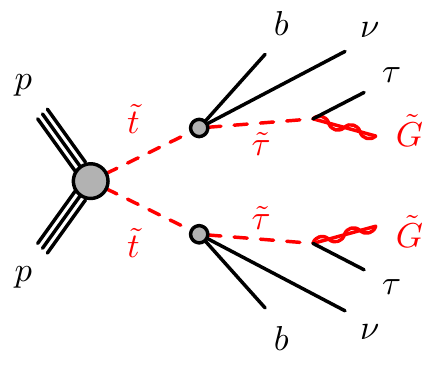}
\hfill
\includegraphics[width=4.2cm]{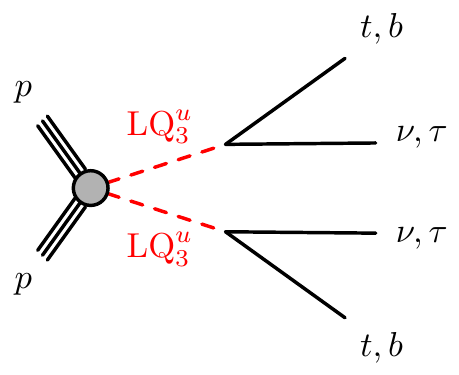}
\hfill
\includegraphics[width=4.2cm]{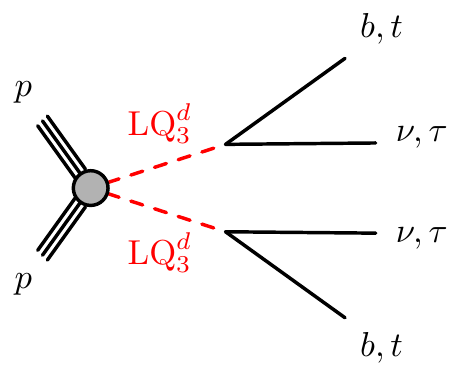}
\hfill
\caption{
Diagrams illustrating the production and decay of particles considered in the simplified models
for the supersymmetric ``stop-stau'' scenario (left) and for scalar leptoquarks of charge $+\frac23e$ (middle) and $-\frac13e$ (right).
}
\label{fig:sim_feynman}
\end{figure}

The investigated SUSY signal model is motivated by gauge-mediated SUSY breaking (GMSB)~\cite{Dine:1981gu,AlvarezGaume:1981wy,Nappi:1982hm}
and natural gauge mediation~\cite{Asano:2010}. 
In this $R$-parity-conserving scenario, only three SUSY particles are assumed to be sufficiently light to be relevant:
the lighter scalar partner of the top quark \stopnoone (top squark or stop),  
the lighter scalar partner of the tau lepton \staunoone (tau slepton or stau), 
and the spin-$3/2$ partner of the graviton, the gravitino \gravitino.
The top squark is assumed to be the lightest squark~\cite{Inoue:1982pi,Ellis:1983ed}
and to be directly pair-produced through the strong interaction.
The gravitino is assumed to be almost massless, making it the lightest SUSY particle (LSP) 
in this scenario.
The search strategy is optimized using a simplified model~\cite{Alwall:2008ve,Alwall:2008ag,Alves:2011wf} with this limited SUSY particle content,
the model parameters being the scalar-fermion masses $\massof{\stopnoone}$ and $\massof{\staunoone}$.
The decay chain is illustrated in the left diagram of \Cref{fig:sim_feynman}:
a three-body decay proceeding through an off-shell chargino
$\stopnoone \rightarrow b \staunoone \nu_{\tau}$
followed by $\staunoone \rightarrow \tau \gravitino$.
This model is referred to as the ``stop-stau'' signal model in the following.
When the LSP is the gravitino,
direct decays of SUSY particles into the gravitino LSP (plus a SM particle)
are very unlikely due to its weak coupling, 
except for the next-to-lightest supersymmetric particle, 
which in $R$-parity-conserving scenarios has no other option than to decay into the gravitino LSP.
Other SUSY models which instead assume the lightest neutralino \ninoone to be the LSP are not studied here,
as this would favor a high branching fraction of $\stopnoone \rightarrow t \ninoone$;
this case has been studied elsewhere by the ATLAS Collaboration~\cite{SUSY-2018-12,SUSY-2018-07,SUSY-2018-08}
and by the CMS Collaboration~\cite{CMS-SUS-19-010,CMS-SUS-19-009,CMS-SUS-19-011,CMS-SUS-20-002}.

Previous searches by the ATLAS Collaboration for signals in this model
used \SI{20}{\ifb} of $\sqrt{s} = \TeVx8$ data taken in Run~1~\cite{SUSY-2014-04}
and \SI{36.1}{\ifb} of \cme data taken in Run~2 of the LHC~\cite{SUSY-2016-19}.
No significant excess was observed in either of these searches,
and thus limits were set on the masses of the top squark and tau slepton. 
These limits exclude top-squark masses of up to \SI{1.16}{\TeV} and tau-slepton masses of up to \SI{1.0}{\TeV} at the \percent{95} confidence level.
The CMS Collaboration has published a related search in a simplified model with pair production of top squarks,
which are also assumed to decay via tau sleptons or tau sneutrinos,
but where the LSP is the lightest neutralino \ninoone instead of the gravitino~\cite{CMS-SUS-19-003}.
This search is based on an integrated luminosity of \ifbx{77.2}
and sets exclusion limits at the \percent{95} confidence level on the top-squark mass of up to \TeVx{1.1} for a nearly massless neutralino.
 
The previous ATLAS Run-2 search in Ref.~\cite{SUSY-2016-19} made use of two event categories:
events where one of the two tau leptons decays leptonically and the other hadronically 
were considered in addition to events where both tau leptons decay hadronically. 
While the branching fractions are almost the same for both categories,
the leptonic decay of the tau lepton yields one neutrino more,
which washes out the kinematic distributions
and on average leads to a lower energy fraction carried by the lepton
compared to the visible decay products from a hadronic tau-lepton decay.
Taken together, the two effects significantly reduce the discriminative power of the selection requirements.
As the sensitivity of the search is thus dominated by the category where both tau leptons decay hadronically,
this \paperornote considers only events with hadronically decaying tau leptons.
These events are separated in two event categories (channels). 
One category selects events with at least two hadronically decaying tau leptons but no lighter leptons,
at least one \bjet and large missing transverse momentum \met (di-tau channel). 
The other category selects events with exactly one hadronically decaying tau lepton,
no electrons or muons, at least two \bjets and large \met (single-tau channel). 
The latter channel extends the sensitivity 
by covering the signal parameter space 
where the tau slepton is relatively light and one of the soft tau leptons easily escapes detection.
Importantly, it also provides good sensitivity to events with pair-produced leptoquarks that decay into third-generation particles,
which correspond to the second benchmark model. 
 
The second benchmark model used in the design of the analysis 
considers pair production of scalar leptoquarks.
It assumes that these only couple to third-generation quark--lepton pairs,
following the minimal Buchmüller--Rückl--Wyler (BRW) model~\cite{Buchmuller:1986zs}.
In addition to the coupling to the third fermion generation that is probed in this analysis,
leptoquarks would need 
to have cross-generational couplings 
in order to explain the anomalies observed in $B$-meson decays.
The search is carried out for both up-type scalar leptoquarks
with fractional charge $Q(\LQu) = +\frac23e$ and decays $\LQu \rightarrow t \nu_{\tau} / b \tau$,
and down-type scalar leptoquarks with $Q(\LQd) = -\frac13e$ and decays $\LQd \rightarrow b \nu_{\tau} / t \tau$.
The production and decay of the leptoquarks are illustrated in \Cref{fig:sim_feynman}.
The model parameters are the leptoquark mass $\massof{\LQud}$ and the branching fraction \BRcharged into a quark and a charged lepton.
For a branching fraction $\BRcharged \sim 0.5$,
most of the decays of the pair of third-generation leptoquarks
yield a final state with one tau lepton, two \bjets and large \met from the tau neutrino.
This signature matches that of the single-tau channel,
which presents unique coverage of leptoquark masses at $\BRcharged \sim 0.5$,
but also has good sensitivity to a wide range of smaller or larger branching fractions.
 
The scalar-\LQ model is the same as was used in a previous ATLAS paper~\cite{EXOT-2017-30}
detailing a search for third-generation leptoquarks based on \SI{36.1}{\ifb} of data taken at \cme.
This earlier paper comprises a dedicated reoptimization of the ATLAS search for pair-produced Higgs bosons and four reinterpretations of ATLAS SUSY searches,
one of which is the previous iteration of the stop-stau search~\cite{SUSY-2016-19}.
Leptoquark masses below at least \TeVx{0.8} are excluded at intermediate values of the branching fraction \BRcharged,
with the lower limit increasing at both small and large \BRcharged,
\eg to $0.96$ \TeVx{(1.02)} at $\BRcharged = 0$~(1) for down-type (up-type) leptoquarks.
Two recent ATLAS searches for top or bottom squark pair production
have been reinterpreted in the same up-type or down-type leptoquark model, respectively~\cite{SUSY-2018-12,SUSY-2018-34}.
Another recent dedicated ATLAS search for pair-produced leptoquarks combines several event categories
which all require at least one hadronically decaying tau lepton plus at least one electron or muon~\cite{EXOT-2019-15}
and are complementary to the final states considered in this \paperornote.
It targets the down-type leptoquark model
and excludes leptoquark masses up to \TeVx{1.43} assuming $\BRcharged = 1$ and up to \TeVx{1.22} assuming $\BRcharged = 0.5$.
The CMS Collaboration has published a search of the full Run-2 dataset for single or pair production of scalar or vector leptoquarks
coupling to third-generation fermions,
considering final-state signatures consisting of a top quark, a tau lepton, a neutrino,
and either no or at least one additional \btagged jet.
This search excludes scalar leptoquarks with masses up to about \TeVx{1.0}~\cite{CMS-EXO-19-015}.
CMS has also reported several searches for third-generation leptoquarks
based on \ifbx{35.9} of Run-2 data~\cite{CMS-SUS-18-001, CMS-B2G-16-028, CMS-B2G-16-027, CMS-EXO-17-016, CMS-EXO-17-029},
which typically set lower limits on the mass of scalar leptoquarks in the range of $0.9$ to \TeVx{1.1}.
 
An additional interpretation of the search results is provided for pair production of vector leptoquarks~\LQv.
Again, it is assumed that the vector leptoquarks can only decay into third-generation SM particles.
The electric charge of the vector leptoquarks and their decay modes
are the same as those of the up-type scalar leptoquarks in the middle diagram of \Cref{fig:sim_feynman}.
The signal selection criteria were not explicitly optimized for this model,
but the kinematic distributions of the decay products are similar for scalar and vector leptoquarks, 
except when the branching fraction of the leptoquarks into a quark and a charged lepton is small,
where tau leptons and \bjets predominantly arise from the leptoquarks decaying into top quarks and neutrinos
rather than directly from the leptoquark decays. 
The signal selection developed for scalar leptoquarks can thus be expected to
also perform very well for the case of vector leptoquarks,
although the relevant energy scales are slightly higher in this case
due to the larger production cross sections at the same mass.
As in the signal model with scalar leptoquarks,
the parameters for the vector-leptoquark model
are the leptoquark mass $\massof{\LQv}$ and the branching fraction \BRchargedv into a quark and a charged lepton.
This is the first time this model is used in a search for leptoquarks by the ATLAS Collaboration.
Models with vector leptoquarks have been considered in several analyses by the CMS Collaboration, including the one in Ref.~\cite{CMS-EXO-19-015},
which excludes vector leptoquarks decaying into $t \bar \nu_\tau$ / $b\tau^+$ with masses up to \TeVx{1.65} 
for pair production in the most favorable coupling scenario.

% End of text imported from the .//tex/introduction.tex input file
\FloatBarrier
 
\section{ATLAS detector}
\label{sec:detector}
% The next lines are included from the .//tex/atlas-detector.tex input file
\newcommand{\AtlasCoordFootnote}{
ATLAS uses a right-handed coordinate system with its origin at the nominal interaction point (IP)
in the center of the detector and the \(z\)-axis along the beam pipe.
The \(x\)-axis points from the IP to the center of the LHC ring,
and the \(y\)-axis points upwards.
Cylindrical coordinates \((r,\phi)\) are used in the transverse plane,
\(\phi\) being the azimuthal angle around the \(z\)-axis.
The pseudorapidity is defined in terms of the polar angle \(\theta\) as \(\eta \equiv -\ln \tan(\theta/2)\)
and is an approximation of the rapidity $y \equiv 0.5\ln\left[ (E+p_z) / (E-p_z) \right]$ in the high-energy limit. 
}

The ATLAS experiment~\cite{PERF-2007-01,ATLAS-TDR-19,PIX-2018-001} 
at the LHC is a multipurpose particle detector
with a forward--backward symmetric cylindrical geometry and a near \(4\pi\) coverage in
solid angle.\footnote{\AtlasCoordFootnote}
It consists of an inner tracking detector surrounded by a thin superconducting solenoid
providing a \SI{2}{\tesla} axial magnetic field, electromagnetic and hadronic calorimeters, and a muon spectrometer.
The inner tracking detector covers the pseudorapidity range \(|\eta| < 2.5\).
It consists of silicon pixel, silicon microstrip, and transition radiation tracking detectors.
Lead/liquid-argon (LAr) sampling calorimeters provide electromagnetic (EM) energy measurements
with high granularity.
A steel/scintillator-tile hadronic calorimeter covers the central pseudorapidity range (\(|\eta| < 1.7\)).
The endcap and forward regions are instrumented with LAr calorimeters
for EM and hadronic energy measurements up to \(|\eta| = 4.9\).
The muon spectrometer surrounds the calorimeters and is based on
three large superconducting air-core toroidal magnets with eight coils each.
The muon spectrometer includes a system of precision tracking chambers and fast detectors for triggering.
A two-level trigger system is used to select events~\cite{TRIG-2019-04}.
The first-level trigger is implemented in hardware and uses a subset of the detector information
to accept events at a rate below \SI{100}{\kHz}.
This is followed by a software-based high-level trigger 
that reduces the accepted event rate to \SI{1}{\kHz} on average
depending on the data-taking conditions.
An extensive software suite~\cite{ATL-SOFT-PUB-2021-001} is used for real and simulated data reconstruction
and analysis, for operation and in the trigger and data acquisition systems of the experiment.

% End of text imported from the .//tex/atlas-detector.tex input file
\FloatBarrier
 
\section{Data and simulated event samples}
\label{sec:samples}
% The next lines are included from the .//tex/samples.tex input file
 
The dataset used in this analysis was collected with the ATLAS detector in proton--proton collisions
provided by the LHC during its second run from 2015 to 2018.
The data was taken at a center-of-mass energy of \cme
with a minimum separation of \unit[25]{ns} between consecutive crossings of proton bunches from the two beams.
Events are selected with triggers on missing transverse momentum~\cite{TRIG-2019-01},
and data-quality requirements are applied to ensure that all elements of the detectors were operational during data-taking~\cite{DAPR-2018-01}.
The total integrated luminosity amounts to \SI{139}{\ifb} with an uncertainty of \percent{1.7}~\cite{ATLAS-CONF-2019-021},
obtained using the LUCID-2 detector~\cite{LUCID2} for the primary luminosity measurements.
 
Monte Carlo (MC) simulation was used to generate samples of collision events,
which model the expected kinematics of the investigated signal and SM background processes.
\Cref{tab:mc_samples} gives a detailed summary of the generation of the different MC samples used in the analysis.
It lists the generators,
the order of the cross-section computation,
the parton distribution function (PDF) sets,
and the sets of tuned parameters (tunes) for the parton shower.
For background processes, the detector response was simulated~\cite{SOFT-2010-01} using the full modeling of the ATLAS detector in $\GEANT$~\cite{Agostinelli:2002hh},
while for the signal samples a faster variant of the simulation was used that relies on a parameterized response of the calorimeters~\cite{ATL-PHYS-PUB-2010-013}.
Except for samples produced with \SHERPA~\cite{Bothmann:2019yzt},
which uses a dedicated parton-shower (PS) modeling and parameter tune developed by the \SHERPA authors,
the parton shower and hadronization simulation for all samples used the A14 tune~\cite{ATL-PHYS-PUB-2014-021},
and the \EVTGEN 
program~\cite{Lange:2001uf} was used to model the decays of $b$- and $c$-hadrons in signal samples and background events.
The effect of multiple concurrent interactions in the same and neighboring bunch crossings (pileup) was modeled
by overlaying the hard-scattering events with simulated inelastic \textit{pp} events generated with \PYTHIAV{8.186}~\cite{Sjostrand:2007gs} using the \NNPDF[2.3lo] set of PDFs~\cite{Ball:2012cx} and the A3 tune~\cite{ATL-PHYS-PUB-2016-017}.  
All simulated events are processed with the same trigger, reconstruction and identification algorithms as the data,
and are weighted to match the observed distribution of the pileup in data.
Dedicated correction factors are applied to simulation to account for differences in efficiencies and energy calibrations
between recorded data and simulations.
In this analysis, data-driven methods are applied
that improve the modeling of the dominant SM background processes by normalizing their contributions to data.
These are described in \cref{sec:background}.
 
\begin{table}[tb]
\caption{
Simulated background and signal samples with the corresponding matrix element and parton shower (PS) generators.
Also, the cross-section order in $\alphas$ used to normalize the event yield
and the parton distribution function (PDF) sets used in the generator and PS simulation are given.
}
\label{tab:mc_samples}
\footnotesize
\centering
\begin{adjustbox}{width=1.\textwidth}
\begin{tabular}{lcccccc}
\toprule
Physics process                     & Generator                                                                             & Parton shower                             & Tune                              & Cross section                                                                 & PDF (generator)                   & PDF (PS) \\
\midrule
\ttbar                              & \POWHEGBOXV{v2}~\cite{Frixione:2007nw,Nason:2004rx,Frixione:2007vw,Alioli:2010xd}     & \PYTHIAV{8.230}~\cite{Sjostrand:2014zea}  & A14~\cite{ATL-PHYS-PUB-2014-021}  & NNLO+NNLL~\cite{Czakon:2011xx}                                                & \NNPDF[3.0nlo]~\cite{Ball:2014uwa}   & \NNPDF[2.3lo]~\cite{Ball:2012cx} \\
Single top                          & \POWHEGBOXV{v2}~\cite{Re:2010bp,Nason:2004rx,Frixione:2007vw,Alioli:2010xd}           & \PYTHIAV{8.230}                           & A14                               & NLO+NNLL~\cite{Aliev:2010zk,Kant:2014oha,Kidonakis:2010ux,Kidonakis:2013zqa}  & \NNPDF[3.0nlo]                       & \NNPDF[2.3lo] \\
$\Vjets~(V=W,Z)$                    & \SHERPAV{2.2.1}~\cite{Bothmann:2019yzt}                                               & \SHERPAV{2.2.1}                           & \SHERPA default                   & NNLO~\cite{Anastasiou:2003ds}                                                 & \NNPDF[3.0nnlo]~\cite{Ball:2014uwa}  & \NNPDF[3.0nnlo] \\
Diboson $VV~(V=W,Z)$                & \SHERPAV{2.2.1} or 2.2.2~\cite{Bothmann:2019yzt}                                      & \SHERPAV{2.2.1} or 2.2.2                  & \SHERPA default                   & NLO~\cite{Gleisberg:2008fv,Schumann:2007mg,ATL-PHYS-PUB-2017-005}             & \NNPDF[3.0nnlo]                      & \NNPDF[3.0nnlo] \\
Triboson $VVV~(V=W,Z)$              & \SHERPAV{2.2.1}                                                                       & \SHERPAV{2.2.1}                           & \SHERPA default                   & NLO~\cite{Gleisberg:2008fv,Schumann:2007mg,ATL-PHYS-PUB-2017-005}             & \NNPDF[3.0nnlo]                      & \NNPDF[3.0nnlo] \\
$\ttV~(V=W,Z)$                      & \MGMCatNLOV{2.3.3}~\cite{Alwall:2014hca}                                              & \PYTHIAV{8.210}~\cite{Sjostrand:2014zea}  & A14                               & NLO~\cite{ATL-PHYS-PUB-2016-005,Alwall:2014hca}                               & \NNPDF[3.0nlo]                       & \NNPDF[2.3lo] \\
$\ttH$                              & \POWHEGBOXV{v2}~\cite{Frixione:2007vw,Alioli:2010xd,Hartanto:2015uka}                 & \PYTHIAV{8.230}                           & A14                               & NLO~\cite{ATL-PHYS-PUB-2016-005,Alwall:2014hca}                               & \NNPDF[3.0nlo]                       & \NNPDF[2.3lo] \\
$\ttbar+WW$                         & \MGMCatNLOV{2.2.2}~\cite{Alwall:2014hca}                                              & \PYTHIAV{8.186}~\cite{Sjostrand:2014zea}  & A14                               & NLO~\cite{Alwall:2014hca}                                                     & \NNPDF[2.3lo]                        & \NNPDF[2.3lo] \\
$\ttbar+WZ$                         & \MGMCatNLOV{2.3.3}                                                                    & \PYTHIAV{8.212}~\cite{Sjostrand:2014zea}  & A14                               & NLO~\cite{Alwall:2014hca}                                                     & \NNPDF[3.0nlo]                       & \NNPDF[2.3lo] \\
$tWZ$                               & \MGMCatNLOV{2.3.3}                                                                    & \PYTHIAV{8.212}                           & A14                               & NLO~\cite{Alwall:2014hca}                                                     & \NNPDF[3.0nlo]                       & \NNPDF[2.3lo] \\
$tZ,\ttbar{t},\ttbar\ttbar$         & \MGMCatNLOV{2.3.3}                                                                    & \PYTHIAV{8.230}                           & A14                               & NLO~\cite{Alwall:2014hca}                                                     & \NNPDF[3.1nlo]~\cite{Ball:2014uwa}   & \NNPDF[2.3lo] \\
\midrule
Stop-stau                           & \MGMCatNLOV{2.6.2}                                                                    & \PYTHIAV{8.212}                           & A14                               & approx. NNLO+NNLL~\cite{Beenakker:2016lwe,Beenakker:1997ut,Beenakker:2010nq,Beenakker:2016gmf} & \NNPDF[2.3lo]  & \NNPDF[2.3lo] \\
Scalar \LQ (\LQud)                  & \MGMCatNLOV{2.6.0}                                                                    & \PYTHIAV{8.230}                           & A14                               & approx. NNLO+NNLL~\cite{Beenakker:2016lwe,Beenakker:1997ut,Beenakker:2010nq,Beenakker:2016gmf} & \NNPDF[3.0nlo] & \NNPDF[2.3lo] \\
Vector \LQ (\LQv)                   & \MGMCatNLOV{2.8.1}                                                                    & \PYTHIAV{8.244}                           & A14                               & LO                                                                                             & \NNPDF[3.0nlo] & \NNPDF[2.3lo] \\
\bottomrule
\end{tabular}
\end{adjustbox}
\end{table}
 
The production of top-quark pairs, with or without an associated Higgs boson, and of single top quarks
in the $s$- or $t$-channel or associated with $W$ bosons was simulated with \POWHEGBOX~\cite{Frixione:2007nw,Nason:2004rx,Frixione:2007vw,Alioli:2010xd},
while associated production of top-quark pairs and a vector boson $V=W$ or $Z$,
as well as top-quark production in other processes (later called ``other top'')
giving smaller contributions ($\ttbar+WW$, $\ttbar+WZ$, $tWZ$, $tZ$, $\ttbar{t}$ and $\ttbar\ttbar$), was simulated with \MGMCatNLO~\cite{Alwall:2014hca}.
The events were interfaced to \PYTHIA~\cite{Sjostrand:2014zea} to model the parton shower, hadronization, and underlying event, using the \NNPDF[2.3lo] set of PDFs~\cite{Ball:2012cx}.
The production of
single vector boson (\Vjets),
diboson ($VV$) and
triboson ($VVV$) events was simulated with \SHERPA using the \NNPDF[3.0nnlo] PDF set~\cite{Ball:2014uwa}.
 
Stop-stau signal samples were produced for various values of $m_{\stop}$ and $m_{\stau}$.
The pair production of top squarks was simulated at leading order with up to two additional partons in \MGMCatNLO.
For the decays of the SUSY particles, the top squark and the tau slepton,
\MADSPIN~\cite{Artoisenet:2012st} was used to preserve spin correlation and finite-width effects.
Both decays are assumed to be prompt, \ie the SUSY particles have a negligible lifetime.
The subsequent decays as well as the hadronization were simulated in \PYTHIA.
Cross sections are calculated including approximate next-to-next-to-leading-order (NNLO) supersymmetric quantum chromodynamics (QCD) corrections,
with resummation of next-to-next-to-leading logarithmic (NNLL) soft gluon terms~\cite{Beenakker:2016lwe,Beenakker:1997ut,Beenakker:2010nq,Beenakker:2016gmf}. 
The matching of matrix element and parton shower was done with the CKKW-L prescription~\cite{Lonnblad:2001iq,Lonnblad:2011xx},
with the matching scale set to one quarter of the top-squark mass.
 
Simulated events with pair production of up- or down-type scalar third-generation leptoquarks \LQud
were generated at next-to-leading order (NLO) in QCD with \MGMCatNLO, 
using the \LQ model of Ref.~\cite{Mandal:2015lca}
that adds parton showers to previous fixed-order NLO QCD calculations~\cite{Kramer:2004df,Kramer:1997hh},
and the \NNPDF[3.0nlo] parton distribution function set with $\alphas(m_Z) = 0.118$. 
\MADSPIN was used for the prompt decays of the leptoquarks into spin-entangled quark--lepton pairs of the third generation.
Parton showering and hadronization were simulated in \PYTHIA with the \NNPDF[2.3lo] PDF set with $\alphas(m_Z) = 0.130$.
The couplings in the Yukawa-type interaction of the leptoquarks with the quark--lepton pair are determined by two parameters:
a common coupling strength $\lambda$ and an additional parameter $\beta$,
with the coupling to a quark and a charged lepton given by $\sqrt{\beta} \lambda$,
and the coupling to a quark and a neutrino by $\sqrt{1 - \beta} \lambda$.
The branching fraction \BRcharged into a quark and a charged lepton is,
except for kinematic effects arising from the mass differences of the decay products, equal to $\beta$.
The leptoquark signal samples were generated for various leptoquark masses $\massof{\LQud}$
and with a fixed parameter value of $\beta = 0.5$, 
so that both decays of the leptoquarks, either into a quark and a neutrino or into a quark and a charged lepton, were possible.
These events can be reweighted to arbitrary branching fractions \BRcharged
to derive the interpretation of the analysis results in the plane of $\massof{\LQud}$ vs.\ \BRcharged.
The coupling parameter $\lambda$ was set to $0.3$,
close to the numeric value of the electromagnetic coupling $e=\sqrt{4\pi\alpha}$, 
resulting in a \LQud width equal to about \percent{0.2} of its mass~\cite{Buchmuller:1986zs,Belyaev:2005ew}.
The cross sections for direct top-squark pair production are used for \LQud pair production,
as both involve massive, scalar, color-charged particles and the production modes are the same.
These cross sections do not include the lepton $t$-channel contributions possible for \LQ pair production,
which are also neglected in Ref.~\cite{Mandal:2015lca} and may lead to corrections at the percent level~\cite{Borschensky:2020hot}.
 
Simulated events with pair production of third-generation vector leptoquarks \LQv
were generated at leading order in QCD with \MGMCatNLO,
using the \LQ model of Ref.~\cite{Baker:2019sli}, 
and the \NNPDF[3.0nlo] parton distribution function set with $\alphas(m_Z) = 0.118$. 
\MADSPIN was used for the prompt decays of the leptoquarks,
and parton showering and hadronization were simulated in \PYTHIA with the \NNPDF[2.3lo] PDF set with $\alphas(m_Z) = 0.130$.
The \LQv in this model corresponds to the $U_1$ state in the BRW classification~\cite{Buchmuller:1986zs}
and carries an electric charge of $Q(\LQv) = +\frac23e$.
The model includes two additional vector states that are needed to obtain a realistic extension of the SM,
a color singlet $Z'$ and a color octet $G'$.
However, the $Z'$ and $G'$ do not appear in the Feynman diagrams considered for pair production of vector leptoquarks, 
as their interactions with the vector leptoquarks are not included in the model.
All $\beta$ parameters are set to zero except for $\beta^{33}_\mathrm{L}$, 
such that only decays to left-chiral fermion fields are allowed,
for which the coupling strength is set to $g_U = 3.0$.
The large value of $g_U$ is motivated by a suppression of the production cross section for additional mediators in a ultraviolet completion of the model,
which might otherwise be in tension with LHC limits
if these mediators are as light as needed to be consistent with the range of \LQ masses considered here. 
As no higher-order computations of the cross sections are available for this vector-leptoquark model,
the leading-order cross sections computed by the event generator are used.
Two different scenarios are considered:
the minimal-coupling scenario with $\kappa_U = \tilde \kappa_U = 1$,
where the \LQ couples to the SM gauge bosons purely through the covariant derivative,
and the Yang--Mills scenario with $\kappa_U = \tilde \kappa_U = 0$,
where the \LQ is a massive gauge boson
and has additional couplings to the SM gauge bosons~\cite{Blumlein:1996qp}.
The two scenarios differ mainly in the pair-production cross section,
which is roughly five times as large in the Yang--Mills scenario at $\massof{\LQv} = \TeVx{1.5}$ as in the minimal-coupling scenario, 
which in turn is roughly four times as large as the pair-production cross section for the scalar \LQud at the same mass. 
 
% End of text imported from the .//tex/samples.tex input file
\FloatBarrier
 
\section{Event reconstruction}
\label{sec:reconstruction}
% The next lines are included from the .//tex/reconstruction.tex input file
 
All events are required to have at least one reconstructed interaction vertex with a minimum of two associated tracks with $\pt > \SI{500}{\MeV}$.
In events with multiple vertices,
the one with the highest sum of squared transverse momenta of associated tracks is chosen as the primary vertex~\cite{ATL-PHYS-PUB-2015-026}.
Events that contain jets that do not satisfy the set of quality criteria described in Ref.~\cite{ATLAS-CONF-2015-029} are rejected
in order to reduce non-collision backgrounds and backgrounds induced by calorimeter noise.
 
Jets are reconstructed from particle-flow objects~\cite{PERF-2015-09} calibrated at the EM scale
using the \antikt algorithm with a radius parameter of $R=0.4$~\cite{Cacciari:2008gp, Fastjet}. 
Since both signal models predict the production of particles with large masses,
only jets in the central region within $|\eta|<2.8$ are used. 
The jets are calibrated following the procedure described in Ref.~\cite{JETM-2018-05} 
and are required to have $\pt > \GeVx{20}$.
To suppress jets from pileup interactions, jet candidates with $\pt < \SI{60}{\GeV}$ and $|\eta|<2.4$ are required to pass the \emph{Tight} working point of the jet vertex tagger~\cite{ATLAS-CONF-2014-018}.  
Selected jets that are likely to originate from the hadronization of a bottom quark are flagged as \bjets
if they lie within $|\eta| < 2.5$ and are tagged by the \texttt{DL1r} algorithm,
a multivariate discriminant based on various inputs such as track impact parameters and displaced secondary vertices~\cite{FTAG-2018-01,ATL-PHYS-PUB-2017-013}.
The \btagging algorithm uses a working point with an efficiency of \percent{77},
with an approximate misidentification probability of
\percent{20} for jets arising from charm quarks,
\percent{6.7} for hadronically decaying $\tau$-leptons, and
\percent{0.9} for light-flavor jets
in simulated \ttbar events.

Tau leptons which decay leptonically are not identified as such,
but are instead reconstructed as a candidate for a prompt electron or muon.
Therefore, in the context of reconstructed analysis objects,
``tau lepton'' will always refer to a hadronic tau lepton,
\ie a tau lepton that decays hadronically.
The visible component of hadronically decaying tau leptons is reconstructed from \antikt jets ($R=0.4$)
built from locally calibrated topological clusters~\cite{PERF-2014-07}, 
with a distance parameter $R = 0.4$ and requiring $\pt > \SI{10}{GeV}$ and $|\eta| < 2.5$~\cite{PERF-2013-06,ATLAS-CONF-2017-029}.
The energy calibration applies a pileup subtraction and a correction to the detector response.
Information from the tracking system improves the energy resolution at low \pt~\cite{ATLAS-CONF-2017-029,PERF-2014-06}.
Tau-lepton candidates are required to have $\pt > \SI{20}{\GeV}$ and lie outside the transition region $1.37<|\eta|<1.52$ between the barrel and endcap calorimeters.
Furthermore, they must have either one or three charged tracks (``prongs'') with a charge sum of $\pm 1$ in units of the elementary charge.
A recurrent neural network algorithm~\cite{ATL-PHYS-PUB-2019-033}
distinguishes hadronically decaying tau leptons from quark- and gluon-initiated jets
by using a combination of high-level discriminating variables as well as tracking and calorimeter measurements.
Its \emph{Medium} working point is used to identify hadronic tau leptons,
with efficiencies of \percent{75} and \percent{60} in simulated Drell--Yan events, 
and background-rejection factors of 35 and 240 in simulated dijet events, 
for one-prong and three-prong decays, respectively.
Electrons misidentified as hadronic tau-lepton candidates are rejected 
using a dedicated boosted decision tree algorithm. 
Reconstructed tau leptons in simulated events are called ``real'' tau leptons
if they can be geometrically matched to a tau lepton in the MC ``truth'' record,
otherwise they are referred to as ``fake'' tau leptons.
 
As described in \cref{sec:selection}, events with prompt electrons or muons are rejected in the analysis selections,
so these only enter in the computation of missing transverse momentum and in the overlap-removal procedure, and are not considered otherwise.
Electron candidates are reconstructed from energy deposits in the electromagnetic calorimeter
that are matched to tracks in the inner detector (ID)~\cite{PERF-2017-01,EGAM-2018-01}. 
They are required to have $\pt > \SI{10}{\GeV}$ and $|\eta| < 2.47$ and pass the \emph{LooseAndBLayer} 
identification requirement.
Muon candidates are reconstructed by combining information from the ID and the muon spectrometer~\cite{MUON-2018-03}.
They are required to have $\pt > \SI{10}{\GeV}$ and $|\eta| < 2.7$ and satisfy the \emph{Medium} identification criteria. 
The absolute value of the longitudinal impact parameter $z_{0}$ of each prompt electron or muon candidate is required to be less than \SI{0.5}{\mm}.
 
An overlap-removal procedure is applied to all selected objects to resolve ambiguities in the reconstruction in several consecutive steps.
First, if two electrons share the same track, the electron with lower transverse momentum is discarded.
Next, tau leptons overlapping with an electron or a muon within $\dRy < 0.2$ are removed,
where the angular distance is measured in units of $\dRy \equiv \sqrt{(\Delta y)^{2} + (\Delta \phi)^{2}}$
with the rapidity $y$ instead of the pseudorapidity $\eta$ to account for cases where particle masses cannot be neglected.
If an electron shares an ID track with a muon, the electron is discarded unless the muon is tagged as a minimum-ionizing particle in the calorimeter,
in which case the muon is discarded.
Jets within $\dRy = 0.2$ of an electron are removed.
In order to suppress electrons from semileptonic heavy-flavor decays,
electrons within $\dRy = 0.4$ of a jet are removed.
Any jet with fewer than three associated tracks is discarded
if a muon is within $\dRy = 0.2$ of the jet or if a muon can be matched to a track associated with the jet.
For the same reason as for electrons, muons within $\dRy = 0.4$ of a jet are removed.
Lastly, jets within $\dRy = 0.4$ of a tau lepton are removed.
 
The missing transverse momentum \vecmet is defined as the negative vector sum of the transverse momenta of all calibrated objects mentioned above,
photons~\cite{EGAM-2018-01}, 
and an additional soft term including all tracks associated with the primary vertex but not matched to any reconstructed object~\cite{PERF-2016-07,ATLAS-CONF-2018-023}.
The magnitude of \vecmet is denoted by \met.
 
% End of text imported from the .//tex/reconstruction.tex input file
\FloatBarrier
 
\section{Event selection}
\label{sec:selection}
% The next lines are included from the .//tex/selection.tex input file
 
The analysis covers two different channels:
the single-tau channel and the di-tau channel.
In both channels, object multiplicities and kinematic variables are used to define several different event selections (analysis regions).
All of these event selections start from a common preselection described below. 
The preselections in the single-tau and di-tau channels
are identical except for the number of tau leptons and \btagged jets.
The sets of events selected in the two channels are thus mutually exclusive and can therefore be statistically combined,
as is done in the interpretation of the results.
 
\subsection{Preselection}
\label{sec:presel}
 
\begin{table}
\caption{
Preselection of the di-tau and single-tau channels.
}
\label{tab:presel_def}
\centering
\begin{tabular}{cc}
\toprule{Di-tau preselection}                 & {Single-tau preselection} \\
\midrule
\multicolumn{2}{c}{\met-trigger fired and $\met > \SI{250}{\GeV}$} \\
\multicolumn{2}{c}{No light leptons ($e$/$\mu$)} \\
\multicolumn{2}{c}{At least two jets} \\
\midrule
At least two hadronic tau leptons     & Exactly one hadronic tau lepton \\
At least one $b$-tagged jet           & At least two $b$-tagged jets \\
\bottomrule
\end{tabular}
\end{table}
 
The preselection requirements for the two channels are summarized in \cref{tab:presel_def}.
Events are selected using an \met trigger~\cite{TRIG-2019-01}.
In combination with the requirement of $\met > \SI{250}{\GeV}$,
this trigger is fully efficient in the phase space that the analysis targets.
As no light leptons are expected from the benchmark signal models
when only hadronically decaying tau leptons are considered, 
events with light leptons are rejected.
Events are required to have at least two jets,
at least one (two) of which must be \btagged in the di-tau (single-tau) channel.
Additionally, events in the di-tau channel are required to have at least two reconstructed tau leptons,
whereas exactly one tau lepton is required in the single-tau channel.
The tight \met and \btag requirements efficiently suppress multijet events
such that their contribution to the analysis regions is negligible. 
This was verified with dedicated data-driven estimates for both channels.

\subsection{Signal regions}
\label{sec:signal_regions}
 
Dedicated signal-enriched regions are defined for each channel,
having been optimized individually by maximizing the estimated discovery significance~\cite{Cowan:2010js}
for benchmark signal models close to the previous exclusion contours. 
The selection requirements for the signal regions are explained in the following,
and a summary is included in the overview of the analysis regions
in \cref{tab:CRVRs_2tau_def} for the di-tau channel
and \cref{tab:CRVRs_1tau_def} for the single-tau channel.
The signal region (SR) in the di-tau channel targets stop-stau signal models with a low to modest mass difference between the top squark and the tau slepton.
This SR is not used for the leptoquark models,
as the final states for that model at $\beta = 0.5$ have only one tau lepton on average.
The case of $\beta = 1.0$, which would yield two tau leptons,
is not within the scope of this \paperornote,
and the requirements on \met and that no leptons be present in the final state
strongly reduce the sensitivity to this scenario.
The single-tau channel employs two signal regions:
a one-bin SR for the model-independent fit,
and a multi-bin SR for the model-dependent fit,
as is discussed in \cref{sec:results}.
Each of the two signal regions in this channel is optimized simultaneously for the scalar-leptoquark signal models
and the stop-stau signal models that have a large mass difference between the top squark and the tau slepton.

\subsubsection*{Di-tau channel}
 
The most discriminating 
variable in the di-tau channel is the ``stransverse'' mass variable~\cite{Lester:1999tx,Lester:2014yga},
which by itself already provides good separation between the signal and the background.
The stransverse mass \mttwo is a generalization of the transverse mass \mt,
which is computed as $
\mt^{2}(\vecpt, \vecmet) = 2 \, \pt \, \met \left(1 - \cos \Delta\phi(\vecpt,\vecmet) \right)
\label{eq:mt}
$ from the transverse momentum of some given particle and the missing transverse momentum.
It generalizes the transverse mass 
for symmetric event topologies where two identical particles each decay into a visible and an invisible product.
In this case the individual transverse momenta of the invisible particles can no longer be directly approximated by the measured missing transverse momentum,
as the information about their individual contributions to the missing transverse momentum is lost.
Using subscripts to refer to the physics objects reconstructed in a collision event in order of decreasing transverse momentum,
for the two leading tau leptons,
\ie the two tau leptons with the largest ($\tau_1$) and second-largest ($\tau_2$) transverse momentum,
$\mttwo(\tau_{1},\tau_{2})$ is computed as
\begin{align*}
\mttwo(\tau_{1},\tau_{2}) =
\min_{\vecqt^{\,a} + \vecqt^{\,b} = \vecmet}
\left(\max\left[\mt( \vecpt^{\tau_1}, \vecqt^{\,a} ), \mt( \vecpt^{\tau_2}, \vecqt^{\,b} ) \right]\right),
\label{eq:mttwo}
\end{align*}
where $a$ and $b$ refer to two invisible particles assumed to be produced with transverse momentum $\vecqt^{\,a,b}$.
The minimum is taken over all possible assignments to $\vecqt^{\,a,b}$ that sum to the measured \vecmet.
The masses of the invisible particles are free parameters and are set to zero.
For the dominant top-quark-related backgrounds, the $\mttwo(\tau_{1},\tau_{2})$ distribution features an endpoint near the $W$-boson mass.
By placing a lower bound at \SI{70}{\GeV} most of this background can be removed,
while efficiently selecting stop-stau signal events,
for which the $\mttwo(\tau_{1},\tau_{2})$ distribution exhibits a tail towards much higher values.
The sensitivity is further enhanced by raising the lower bound on \met to \SI{280}{\GeV}
and requiring the two leading tau leptons to carry electric charges with opposite signs,
a criterion later denoted by $OS(\tau_{1},\tau_{2}) = 1$.

\subsubsection*{Single-tau channel}
 
Both signal regions in the single-tau channel have a lower bound on \met at \SI{280}{\GeV}
and on the sum of the transverse masses of the \bjets,
$\mtsumbb = \mt(b_{1}) + \mt(b_{2})$, at \SI{700}{\GeV}.
In this expression and the following, $\mt(A)$ for a given particle $A$ should be read as $\mt(A) \equiv \mt(\vecpt^A,\vecmet)$.
The one-bin SR requires $\mt(\tau) > \SI{300}{\GeV}$ and $\st > \SI{800}{\GeV}$,
where $\st$ is defined as the scalar sum of the transverse momenta of the tau lepton and the two leading jets,
$\st = \pt(\tau) + \pt(\text{jet}_{1}) + \pt(\text{jet}_{2})$.
While both the stop-stau and \LQud signals show fairly similar behavior in most kinematic variables, 
their $\pt(\tau)$ distributions differ. 
This is due to the large mass difference in the stop-stau target scenario,
so that the tau leptons are softer than those produced in the \LQud decay.
To account for the different shapes of the transverse momentum distributions of the tau leptons,
the second SR is defined with three bins in $\pt(\tau)$.
The first two $\pt(\tau)$ bins cover 50 to \GeVx{100} 
and 100 to \GeVx{200},
and the last bin all values beyond \SI{200}{\GeV}.
To reduce the statistical uncertainty in the three $\pt(\tau)$ bins,
two selection requirements are loosened relative to the one-bin SR:
the minimum $\mt(\tau)$ requirement is lowered to \SI{150}{\GeV},
and the minimum $\st$ requirement to \SI{600}{\GeV}.
As the one-bin SR is a subset of the multi-bin SR,
they cannot be combined in the statistical interpretation of the results discussed in \cref{sec:results}.
A multi-bin SR based on \st instead of $\pt(\tau)$ was also tested, 
but was found to have lower sensitivity.

% End of text imported from the .//tex/selection.tex input file
\FloatBarrier
 
\section{Background estimation}
\label{sec:background}
% The next lines are included from the .//tex/background.tex input file
 
The background in the signal regions is dominated by \ttbar and single-top production,
which can yield events with a final state similar to the signal processes.
Dedicated control regions are defined for these background processes.
Top-quark production can contribute to the background in different ways.
Events with \ttbar production,
where both $W$~bosons arising from the top-quark decay into a hadronic tau lepton,
have two real tau leptons.
This process, denoted by \ttTwoReal, 
contributes to the di-tau channel if both hadronic tau leptons are correctly identified.
If instead only one of the $W$~bosons from the \ttbar system gives a hadronic tau lepton
which is correctly identified,
and the second $W$~boson decays hadronically,
the resulting jet from the second $W$-boson decay can be misidentified as a tau lepton,
and such an event can then still satisfy the di-tau channel selection criteria.
While the misidentification probability is of the order of a few percent,
the larger branching fraction of hadronic $W$~decays
and the less pronounced endpoint in the $\mttwo(\tau_{1}, \tau_{2})$ distribution
for \ttbar events with one real and one fake tau lepton still leads to a significant contribution in the di-tau channel.
This type of event can also enter the single-tau channel selection,
if the jet from the second $W$~boson is not misidentified as a tau lepton.
Di-tau \ttbar events in which only one of the two identified tau leptons is real,
and single-tau \ttbar events with one real tau lepton, are referred to as \ttOneReal events.
Lastly, fully hadronic \ttbar decays, without any real tau leptons that pass the selections
in either the single-tau or di-tau channel, are referred to as \ttFake events.
Due to their different kinematics, the simulated \ttbar events are separated into these three event types,
\ttTwoReal, \ttOneReal, and \ttFake,
and treated as separate background components in the following.
 
Subdominant contributions to the SM background arise from singly produced vector bosons (\Wjets and \Zjets events)
and production of vector bosons in association with top-quark pairs (\ttV).
In addition, multiboson production, \ttbar production in association with a Higgs boson (\ttH) and other top-related processes yield small contributions.
These subdominant processes are normalized according to the theory cross-section predictions and the integrated luminosity measured in data.
 
The normalization factors for the MC predictions for \ttbar and single-top production
are extracted in a simultaneous binned maximum-likelihood fit to the observed data in the control regions (CRs).
This fit, where no signal contributions are included,
is referred to as the background-only fit. 
The CRs are designed to be enriched in a given background process and to be kinematically as similar to the SRs as possible,
while maintaining sufficient purity and a high enough event yield with negligible contamination from signal.
In addition to the data yields in the CRs,
the expected yields and statistical and systematic uncertainties from MC simulation,
described in \cref{sec:systematics},
are input to the background-only fit.
The yields obtained from the background-only fit can then be extrapolated to dedicated validation regions (VRs)
to assess the accuracy of the background estimate.
All CR, VR and SR selections are mutually exclusive so that they are statistically independent as required for the fit.
The CR and VR selections are introduced in \cref{sec:bkg_est_2tau,sec:bkg_est_1tau} for the di-tau and single-tau channel, respectively.
In addition to the background-only fit,
model-dependent and model-independent fit setups~\cite{Baak:2014wma}
which include the signal regions
are used for the interpretation of the results as further detailed in \cref{sec:results}.

\subsection{Di-tau channel}
\label{sec:bkg_est_2tau}
 
\begin{table}
\newcommand{\emptycell}{---} 
\caption{
Definitions of the \ttbar control and validation regions and the signal region in the di-tau channel.
A dash signifies that no requirement on the given variable is applied,
while brackets indicate an allowed range for the variable.
These requirements extend those of the di-tau preselection from \cref{tab:presel_def}.
}
\label{tab:CRVRs_2tau_def}
\sisetup{parse-numbers=false}
\centering
\begin{adjustbox}{width=1.\textwidth}
\begin{tabular}{lccccc}
\toprule{Variable}                  & {CR \ttTwoReal}           & {CR \ttOneReal}           & {VR \ttTwoReal}           & {VR \ttOneReal}           & {SR} \\
\midrule
$\met$                      & \emptycell                & \emptycell                & \emptycell                & \emptycell                & $>\SI{280}{\GeV}$ \\
$OS(\tau_{1},\tau_{2})$     & 1                         & \emptycell                & 1                         & \emptycell                & 1                 \\
$\mttwo(\tau_{1},\tau_{2})$ & $<\SI{35}{\GeV}$          & $<\SI{35}{\GeV}$          & $\SI{[35,70]}{\GeV}$      & $\SI{[35,70]}{\GeV}$      & ~~$>\SI{70}{\GeV}$  \\
$\mtautau$                  & $>\SI{50}{\GeV}$          & $>\SI{50}{\GeV}$          & \emptycell                & \emptycell                & \emptycell        \\
$\mt(\tau_{1})$             & $>\SI{50}{\GeV}$          & $<\SI{50}{\GeV}$          & $>\SI{70}{\GeV}$          & $<\SI{70}{\GeV}$          & \emptycell        \\
\bottomrule
\end{tabular}
\end{adjustbox}
\end{table}
 
\begin{figure}
\centering
\includegraphics[width=12cm]{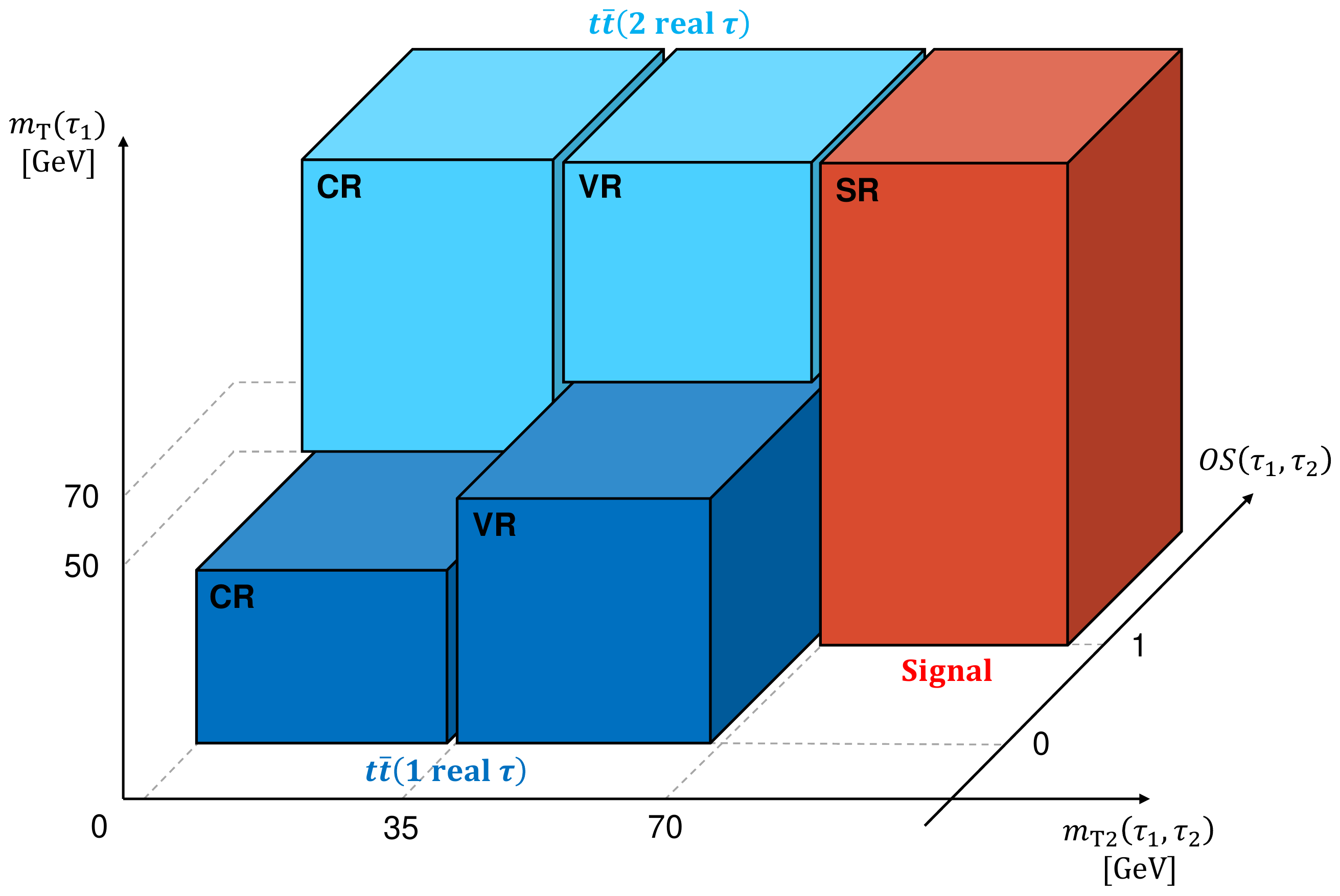}
\caption{
Overview of the selections defining the control, validation and signal regions in the di-tau channel in the phase-space spanned
by the variables
$\mttwo(\tau_1, \tau_2)$,
$\mT(\tau_1)$, 
and $OS(\tau_1,\tau_2)$,
where $OS(\tau_1,\tau_2) = 1$ means that the reconstructed charges of the two leading tau leptons have opposite signs.
In addition to these variables,
$\met > \GeVx{280}$ is required for the signal region,
and $\mtautau > \GeVx{50}$ for the control regions.
The complete definitions are summarized in \cref{tab:CRVRs_2tau_def}.
}
\label{fig:selections_2tau}
\end{figure}
 
\Cref{tab:CRVRs_2tau_def} summarizes the selections that define the control, validation and signal regions in the di-tau channel.
One control region and one validation region are defined in this channel
for each of the \ttTwoReal and \ttOneReal processes.
Their location in the phase-space spanned by
$\mttwo(\tau_{1},\tau_{2})$, $\mt(\tau_{1})$, and $OS(\tau_{1},\tau_{2})$ is illustrated in \Cref{fig:selections_2tau}.
The CRs and VRs sit in the $\mttwo(\tau_{1}, \tau_{2})$ sideband below \SI{70}{\GeV},
above which the SR is located, and are separated at \SI{35}{\GeV}.
 
Top-quark pair-production events in which only one of the $W$~bosons decays leptonically,
with one real tau lepton and one fake tau lepton, 
feature an endpoint in the \mt distribution of the real tau lepton near the $W$~mass.
The reason is that the dominant source of \met is the tau neutrino from the $W$~decay.
By contrast, for \ttbar events with two real tau leptons, 
two tau neutrinos contribute to the \met and there is no distinct endpoint in \mt.
This difference in the shapes of the \mt distributions is exploited in the selection of \ttTwoReal and \ttOneReal events.
In the majority of \ttOneReal events in the di-tau channel,
the real tau lepton corresponds to the leading reconstructed tau lepton.
A requirement on $\mt(\tau_{1})$ at 50~\GeVx{(70)} is thus used to separate the \ttTwoReal CR (VR) from the \ttOneReal CR (VR).
By requiring the leading and subleading tau lepton in the \ttTwoReal CR and VR selections
to carry electric charges of opposite sign, $OS(\tau_{1},\tau_{2}) = 1$, the purity is further increased.
In addition, a lower bound on the invariant mass of the two tau leptons computed from the visible decay products, \mtautau, at \SI{50}{\GeV} is applied
to reduce the contribution from \Zjets events.
 
Distributions of the main discriminating variables $\mttwo(\tau_{1}, \tau_{2})$ and \met
in the control and validation regions of the di-tau channel
are shown in \Cref{fig:CRVRs_2tau_dataMC}.
\MCdatascaled
Their values are given in \cref{sec:results}.
\MCdatagood
 
\begin{figure}
\centering
\includegraphics[width=0.49\linewidth]{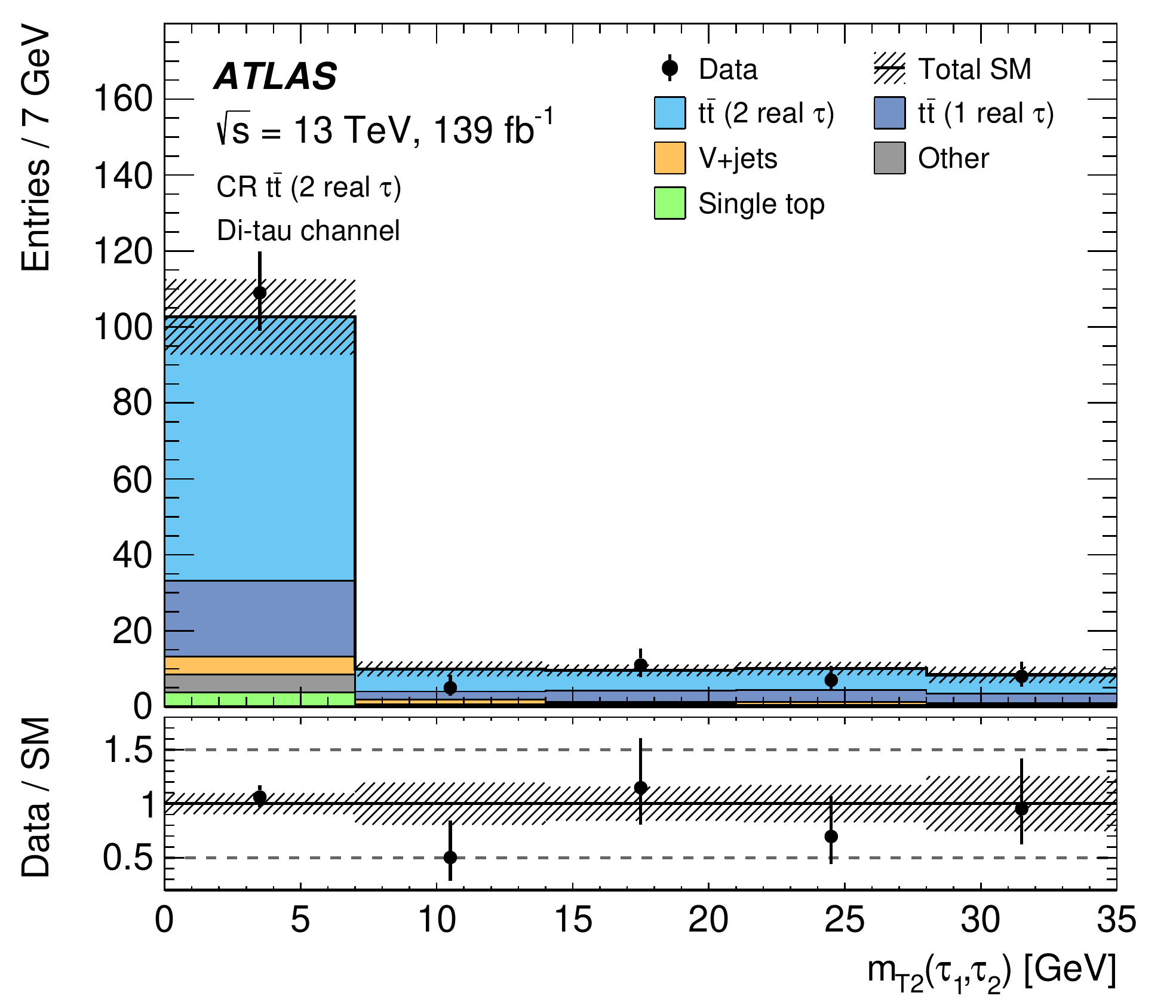}
\includegraphics[width=0.49\linewidth]{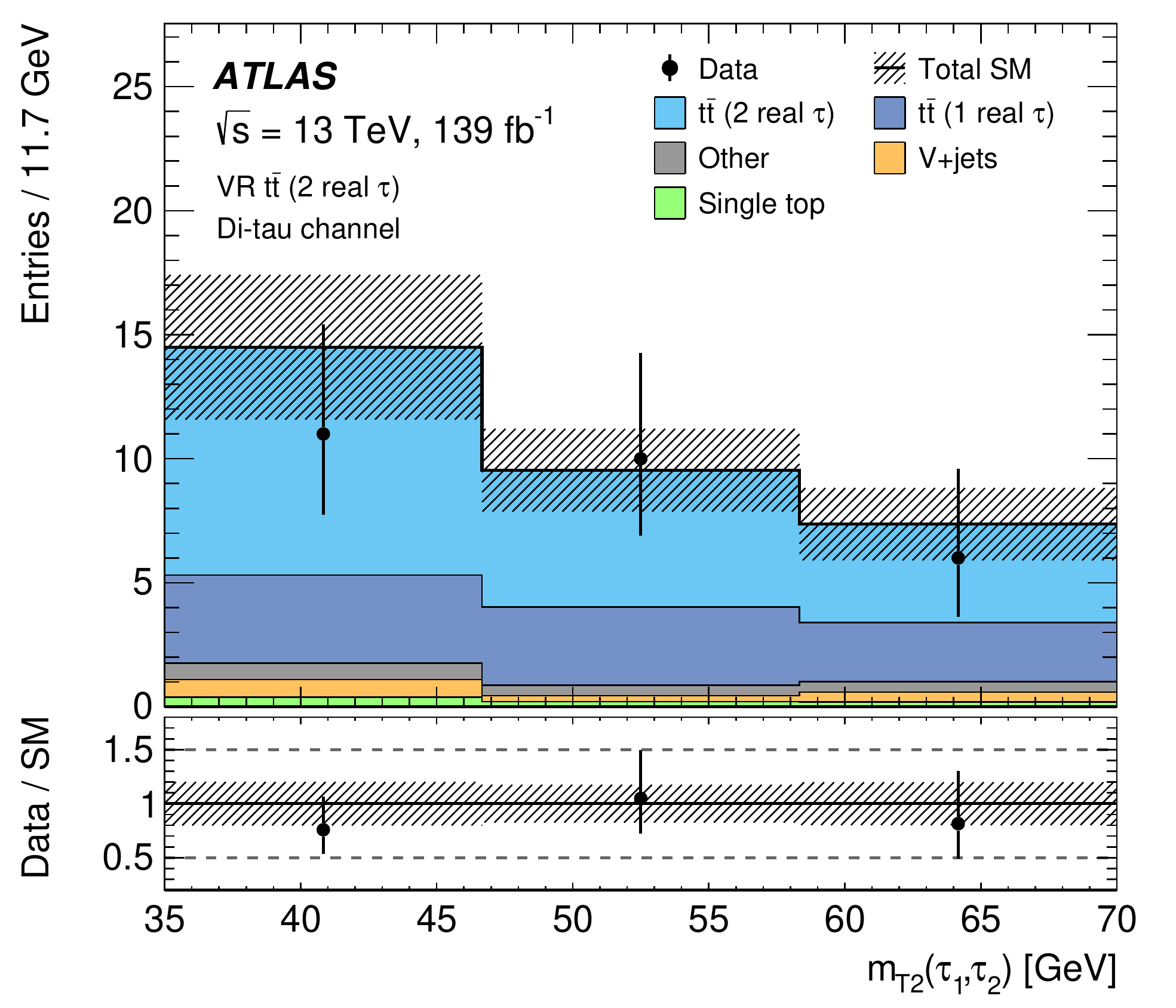}
\includegraphics[width=0.49\linewidth]{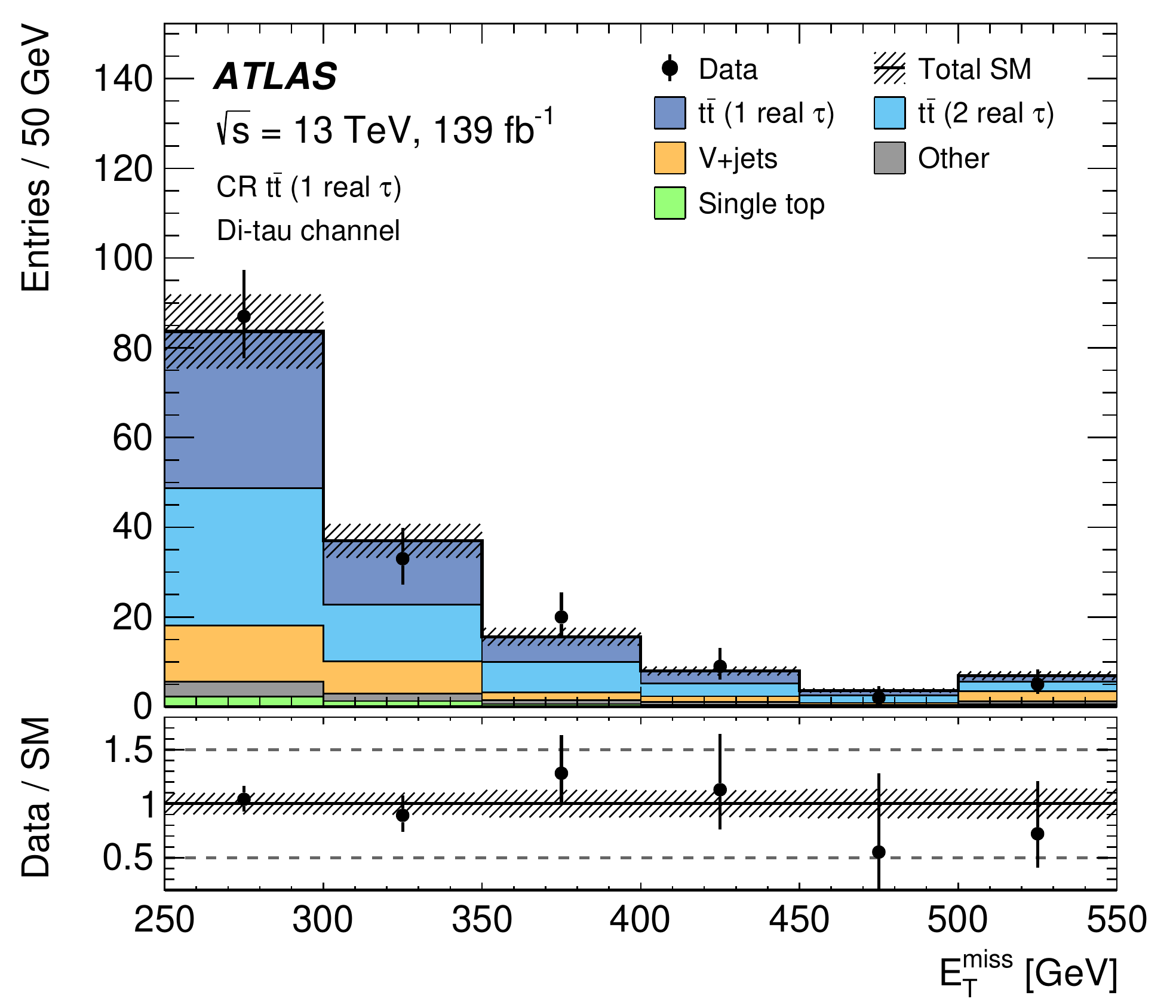}
\includegraphics[width=0.49\linewidth]{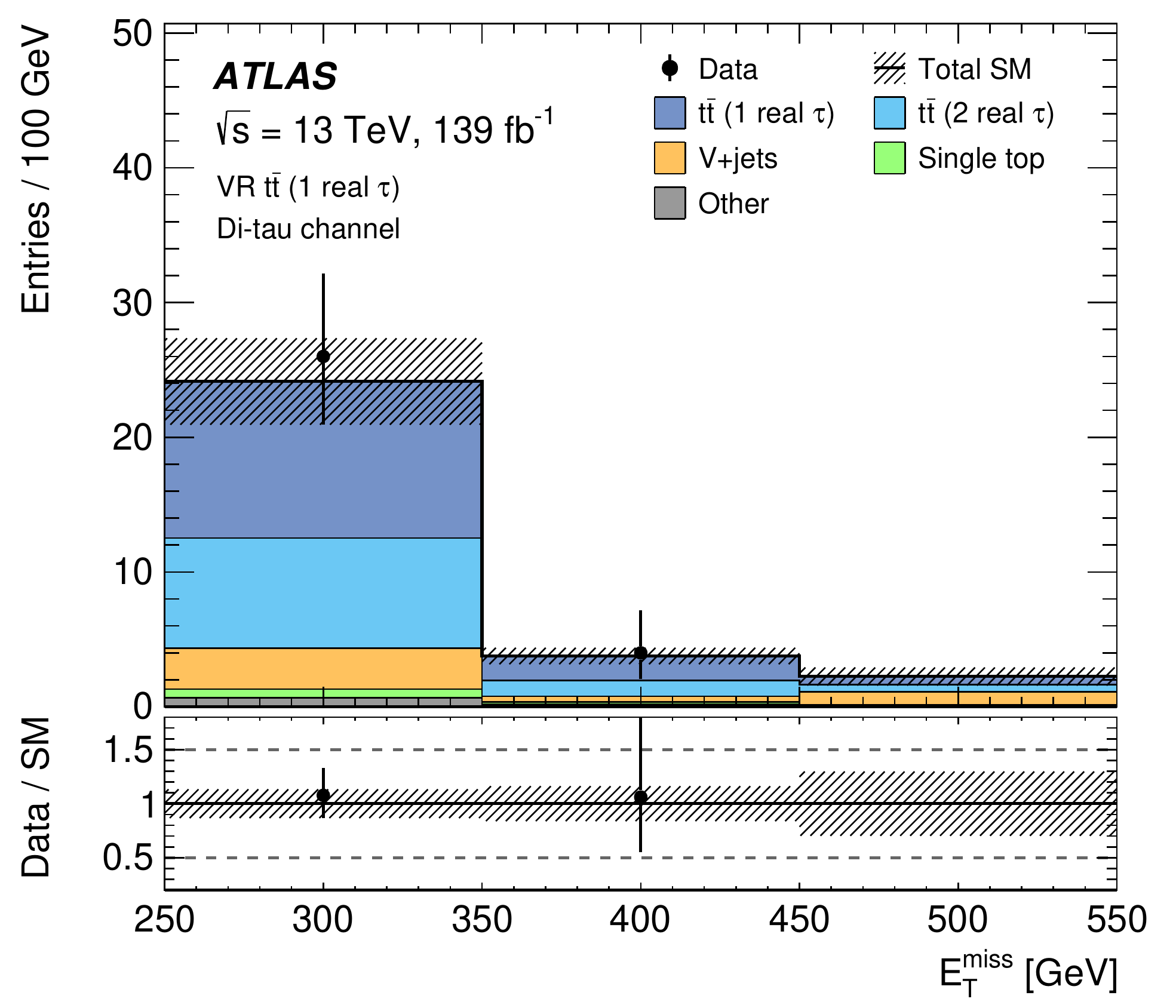}
\caption{
Distributions of $\mttwo(\tau_{1}, \tau_{2})$ and \met in the di-tau channel.
The left-hand plots show the control regions and the right-hand plots the validation regions,
with $\mttwo(\tau_{1}, \tau_{2})$ in the \ttTwoReal CR and VR in the top row and
\met in the \ttOneReal CR and VR in the bottom row.
The CRs and VRs have different requirements on the transverse mass $\mt(\tau_{1})$.
\StackedHisto
\BandIsStatSyst
\Postfit
\BackgroundOtherInPlots{\ttFake, \ttX, multiboson, and other top} 
\RightBinOverflow
}
\label{fig:CRVRs_2tau_dataMC}
\end{figure}

\FloatBarrier

\subsection{Single-tau channel}
\label{sec:bkg_est_1tau}
 
\begin{figure}
\centering
\includegraphics[width=0.65\linewidth]{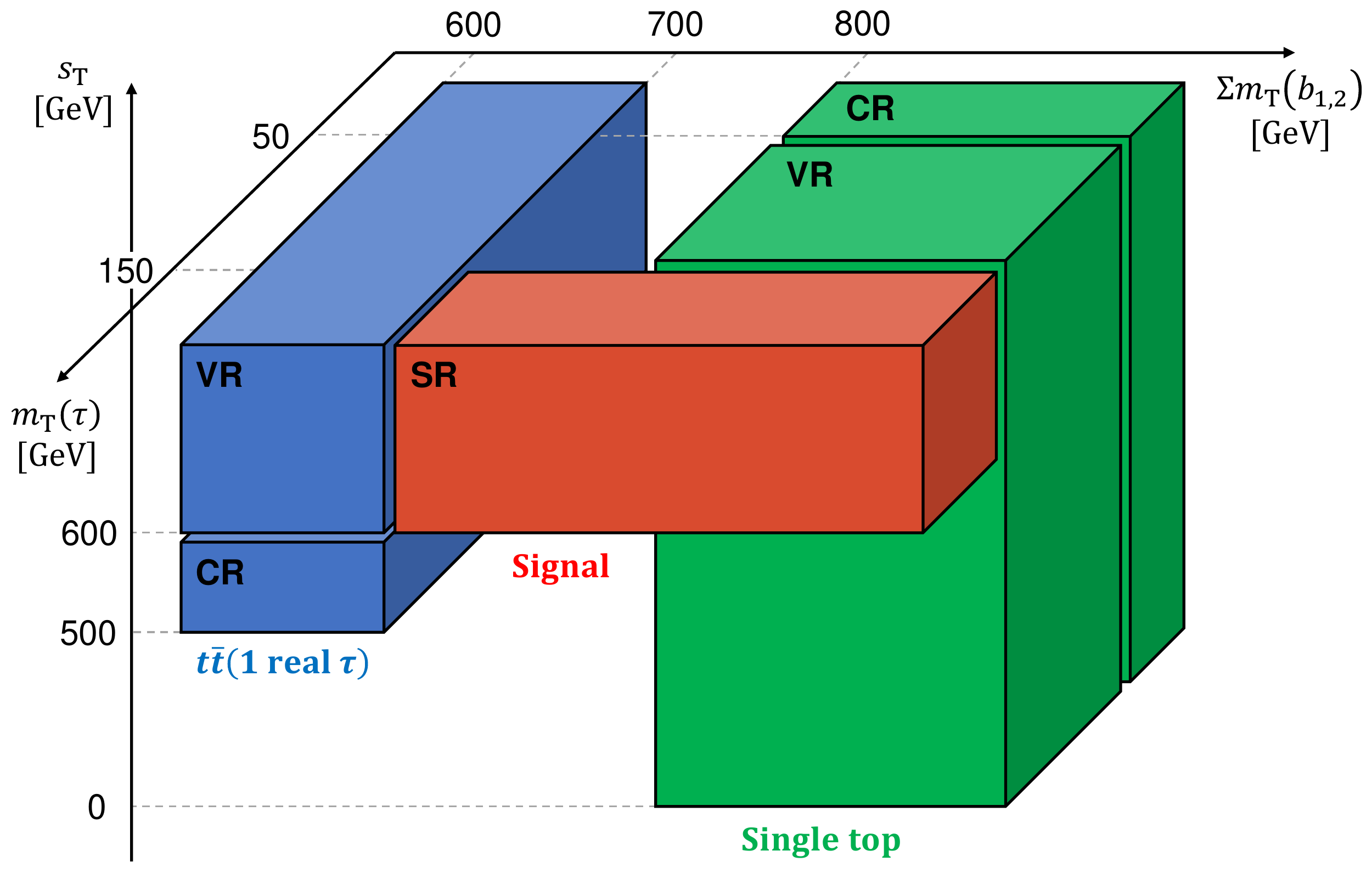}
\caption{
Overview of the selections defining the control and validation regions and the multi-bin signal region in the single-tau channel
in the phase-space spanned by the variables \mtsumbb, $\mt(\tau)$ and \st.
In addition to these variables,
$\met > \GeVx{280}$ is required for the signal region,
and $\pt(\tau_1) > \GeVx{80}$ for the single-top control and validation regions.
The complete definitions are summarized in \cref{tab:CRVRs_1tau_def}.
}
\label{fig:selections_1tau}
\end{figure}
 
\begin{table}
\newcommand{\emptycell}{---\xspace} 
\caption{
Definitions of the \ttOneReal and single-top control and validation regions and the signal region in the single-tau channel.
A dash signifies that no requirement on the given variable is applied,
while brackets indicate an allowed range for the variable. 
In the last column, parentheses enclose the values and ranges used for the multi-bin SR. 
The binning in $\pt(\tau)$ of the multi-bin SR, abbreviated with ``binned'',
is $[50, 100]$, $[100,200]$, and $>\GeVx{200}$.
These requirements extend those of the single-tau preselection from \cref{tab:presel_def}.
}
\label{tab:CRVRs_1tau_def}
\sisetup{parse-numbers=false}
\centering
\begin{adjustbox}{width=1.\textwidth}
\begin{tabular}{lccccc}
\toprule{Variable}                      & {CR \ttOneReal}               & {CR single top}               & {VR \ttOneReal}             & {VR single top}               & {SR} \\
\midrule
$\met$                          & $>\SI{280}{\GeV}$             & $>\SI{280}{\GeV}$             & $>\SI{280}{\GeV}$           & $>\SI{280}{\GeV}$             & $>\SI{280}{\GeV}$        \\
$\st$                           & $\SI{[500,600]}{\GeV}$        & \emptycell                    & $>\SI{600}{\GeV}$           & \emptycell                    & $>\SI{800 (600)}{\GeV}$  \\
$\mtsumbb$                      & $\SI{[600,700]}{\GeV}$        & $>\SI{800}{\GeV}$             & $\SI{[600,700]}{\GeV}$      & $>\SI{800}{\GeV}$             & $>\SI{700}{\GeV}$        \\
$\mt(\tau)$                     & \emptycell                    & ~~$<\SI{50}{\GeV}$            & \emptycell                  & $\SI{[50,150]}{\GeV}$         & $>\SI{300 (150)}{\GeV}$  \\
$\pt(\tau)$                     & \emptycell                    & ~~$>\SI{80}{\GeV}$            & \emptycell                  & ~~$>\SI{80}{\GeV}$            & \emptycell (binned)      \\
\bottomrule
\end{tabular}
\end{adjustbox}
\end{table}
 
For the two dominant processes in the single-tau channel, \ttbar production with one real tau lepton and single-top production,
again two pairs of control and validation regions are defined.
The definitions are illustrated in \Cref{fig:selections_1tau} and summarized in \cref{tab:CRVRs_1tau_def}.
In contrast to the di-tau CRs and VRs,
the larger available number of events in the single-tau channel
allows the lower bound on \met used in the CR and VR selections to be the same as for the SR.
The \ttOneReal control and validation regions in the single-tau channel
are placed in the \mtsumbb sideband between $600$ and $\SI{700}{\GeV}$.
The control region is located in the \st window from $500$ to $\SI{600}{\GeV}$,
and the validation region covers the range above $\SI{600}{\GeV}$.
The normalization of the \ttOneReal process is obtained from a simultaneous fit of both control regions for this process,
one from each channel.
The CR and VR for the single-top background sit in the $\mt(\tau)$ window from 0 to \SI{50}{\GeV} and $50$ to \SI{150}{\GeV}, respectively.
Additionally,
lower bounds on \mtsumbb at \SI{800}{\GeV} and
on $\pt(\tau)$ at \SI{80}{\GeV}
increase the purity of both the single-top CR and VR.
Events from \ttOneReal are less likely to fulfill the $\pt(\tau)$ requirement,
which favors high-energy decay products.
They also tend to have lower \mtsumbb,
as the transverse mass computed for the sub-leading \bjet has a quite distinct endpoint near the top-quark mass.
 
\Cref{fig:CRVRs_1tau_dataMC} shows the distribution of \st and $\mt(\tau)$
in the \ttOneReal CR and VR and
in the single-top CR and VR of the single-tau channel.
\MCdatascaled
These are consistent with one for the \ttTwoReal and \ttOneReal backgrounds,
but much smaller than one for the single-top background as discussed further in \cref{sec:results}.
Therefore, the contribution of scaled single-top events to the single-top CR and VR in the figure is very low,
whereas it is \percent{43} before applying the normalization factors.
\MCdatagood

\begin{figure}
\centering
\includegraphics[width=0.49\linewidth]{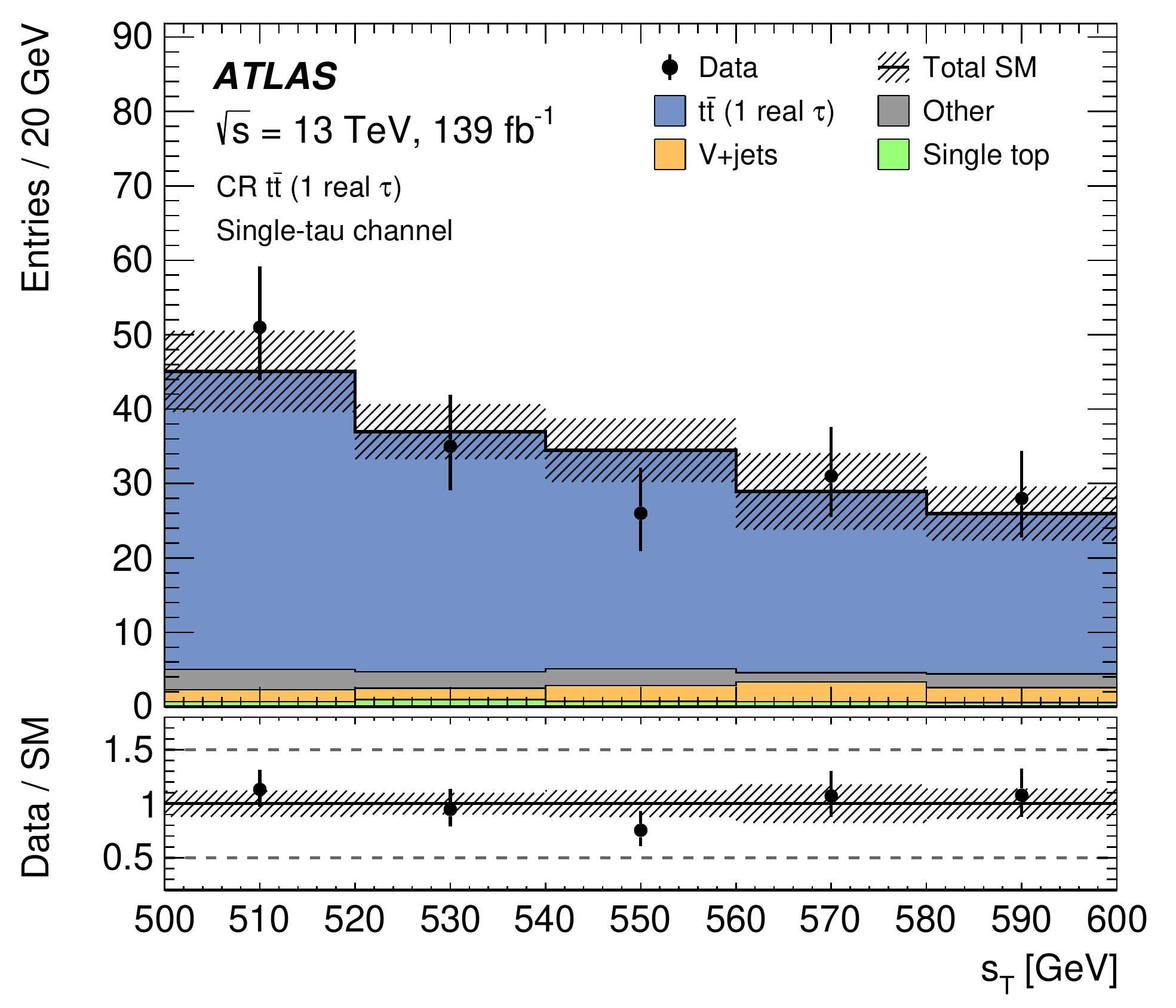}
\includegraphics[width=0.49\linewidth]{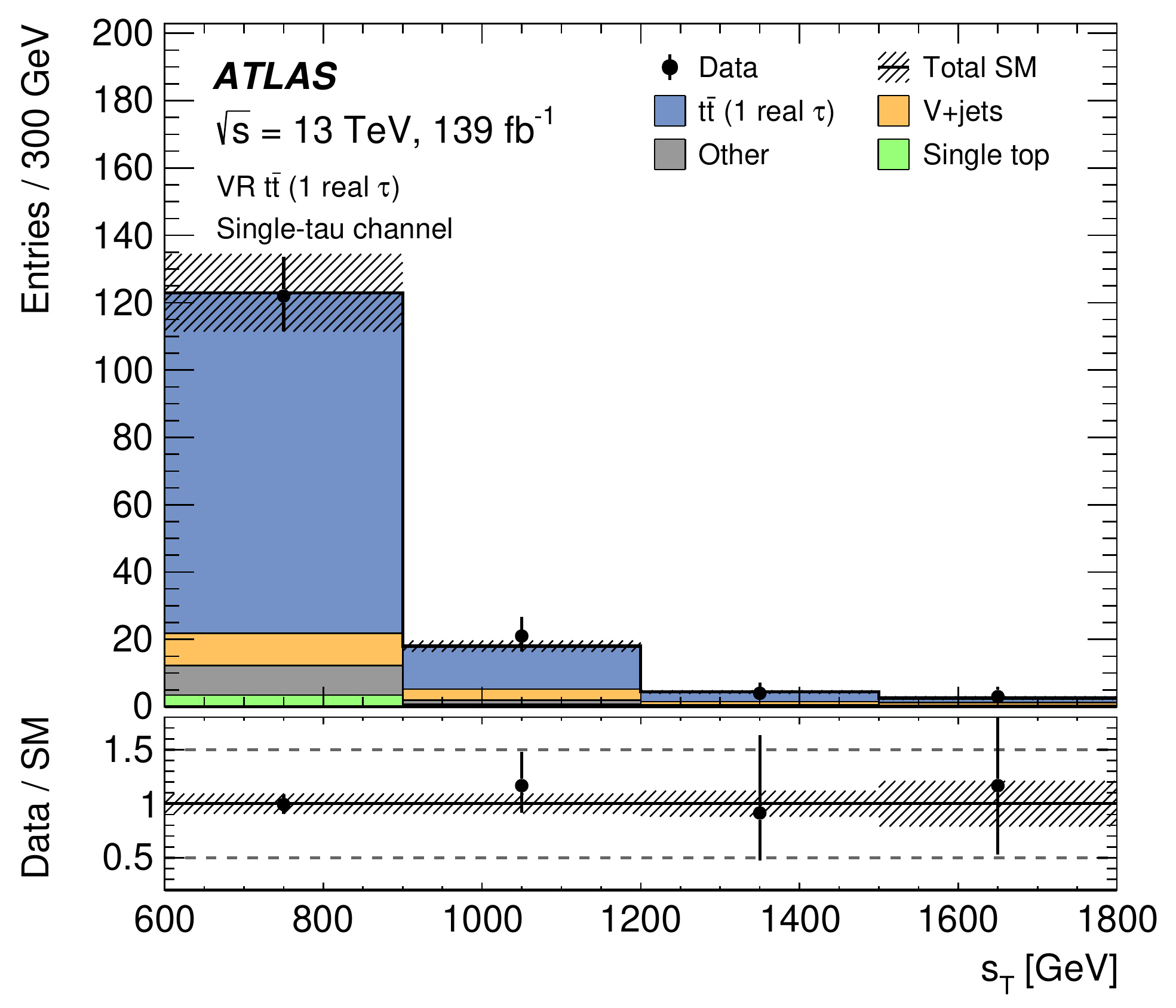}
\includegraphics[width=0.49\linewidth]{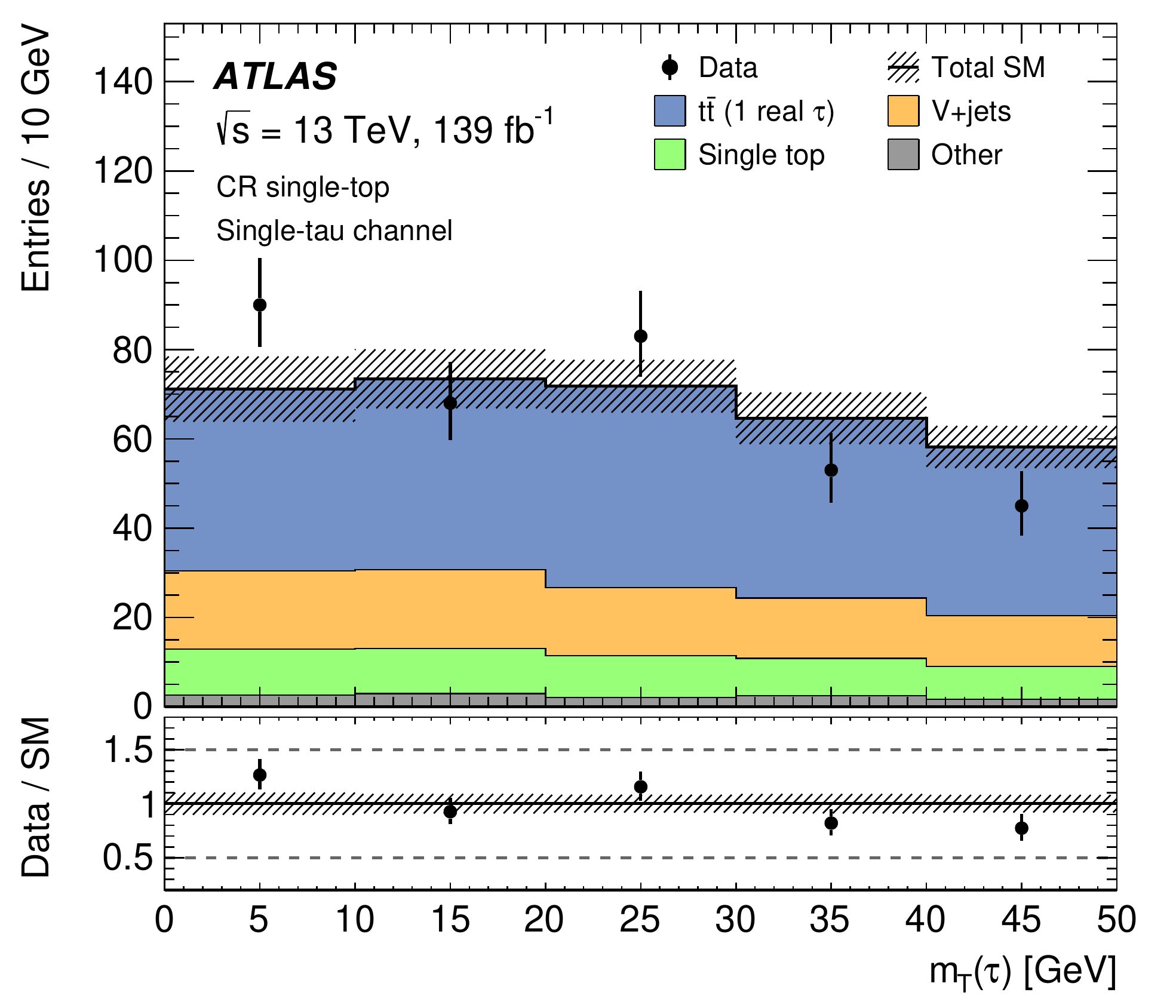}
\includegraphics[width=0.49\linewidth]{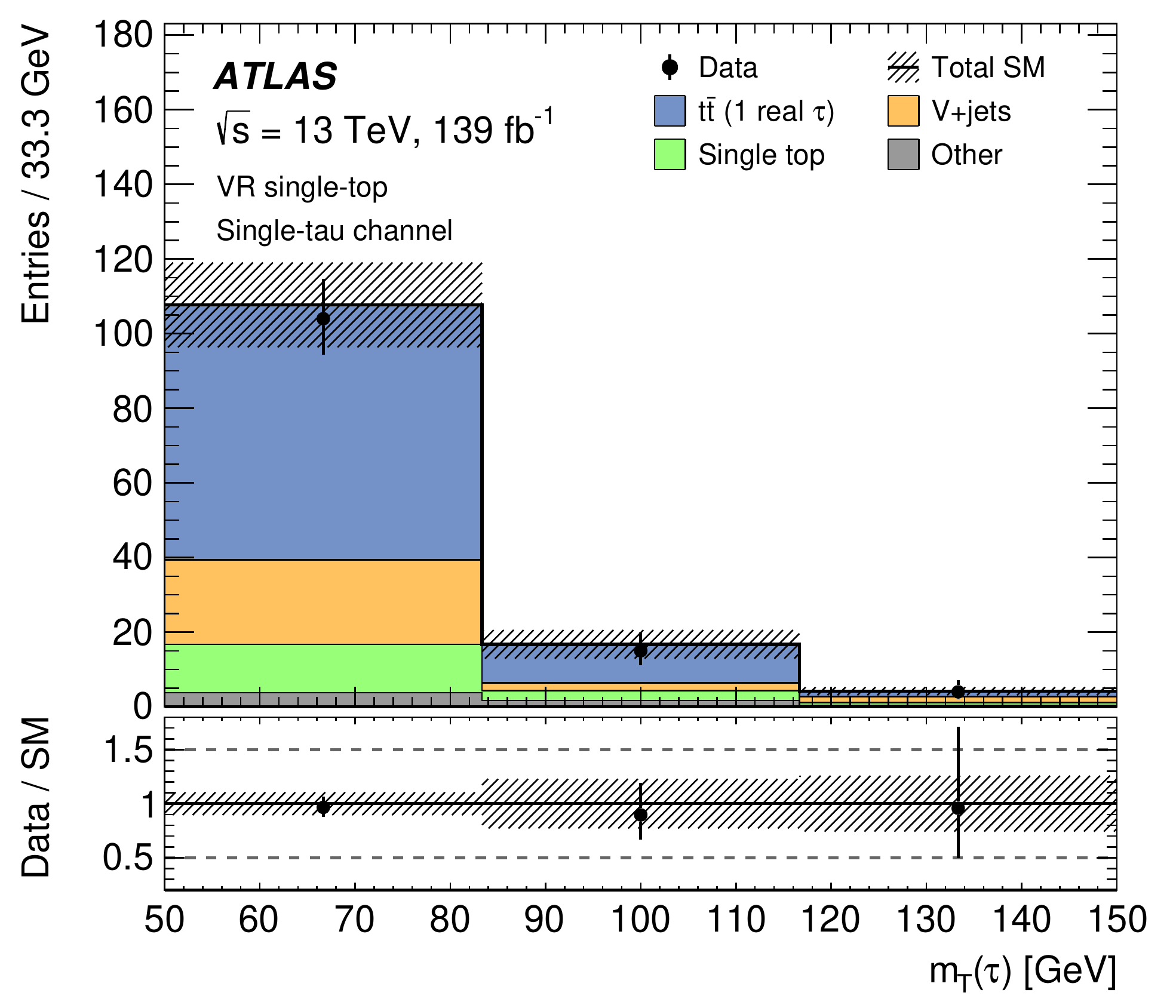}
\caption{
Distributions of \st and $\mt(\tau)$ in the control regions and the validation regions of the single-tau channel.
The top row shows \st
in the \ttOneReal CR (left) and \ttOneReal VR (right),
the bottom row shows $\mt(\tau)$
in the single-top CR (left) and single-top VR (right).
\StackedHisto
\BandIsStatSyst
\PostfitSingletau
\BackgroundOtherInPlots{\ttFake, \ttX, multiboson, and other top} 
\RightBinOverflowC{ in the upper right plot}
}
\label{fig:CRVRs_1tau_dataMC}
\end{figure}
 
% End of text imported from the .//tex/background.tex input file
\FloatBarrier
 
\section{Systematic uncertainties}
\label{sec:systematics}
% The next lines are included from the .//tex/systematics.tex input file
 
The expected yields for signal and background processes are subject to experimental and theoretical systematic uncertainties.
These uncertainties are implemented as variations
which are parameterized as functions of nuisance parameters with Gaussian probability densities in the likelihood fits.
 
Experimental uncertainties comprise systematic uncertainties in the reconstruction, identification, calibration and corrections performed for the physics objects used in the analysis.
Energy resolution and calibration uncertainties apply to all objects.
For tau leptons, additional experimental systematic uncertainties arise from the reconstruction and identification efficiencies.
Since events with prompt electrons and muons are rejected at preselection level,
the related uncertainties in the reconstruction and identification are negligible in the analysis regions. 
For jets, additional uncertainties from the pileup subtraction, pseudorapidity intercalibration, flavor composition, and punch-through effects,
as well as uncertainties in the flavor-tagging and jet-vertex tagging efficiencies, are considered.
Systematic uncertainties affecting the energy or momentum of calibrated objects are propagated to the \met calculation,
and an additional uncertainty due to the contribution of the soft-track term is considered.
To test the robustness of the analysis against a potential mismodeling of events with two fake tau leptons,
it was verified that an additional uncertainty of \percent{100} in the \ttFake background
leads to a negligible decrease in the exclusion reach for the stop-stau signal model.
Common sources of experimental uncertainty are assumed to be correlated across all regions and between the background processes and the signal.
 
\begin{table}[tb]
\caption{
Relative systematic uncertainties in the estimated number of background events in the signal regions.
In the lower part of the table, a breakdown of the total uncertainty into different categories is given.
For the multi-bin SR, the breakdown refers to the integral over all three $\pt(\tau)$ bins.
\UncertaintiesDoNotSumUp
}
\label{tab:SRs_unc_breakdown}
\centering
\small
\begin{tabular}{lccc}
\toprule
Systematic uncertainty        & Di-tau SR             & Single-tau one-bin SR           & Single-tau multi-bin SR  \\
\midrule
Total                         & \percent{25}~~        & \percent{17}~~                  & \percent{17}~~    \\
\midrule
Jet-related                   & \percent{19}~~        & ~~~\percent{4.2}                & ~~~\percent{3.9}  \\
Tau-related                   & ~~~\percent{4.7}      & ~~~\percent{5.5}                & ~~~\percent{4.3}  \\
Other experimental            & ~~~\percent{3.7}      & ~~~\percent{1.0}                & ~~~\percent{0.8}  \\
Theoretical modeling          & \percent{13}~~        & \percent{17}~~                  & \percent{19}~~    \\
MC statistics                 & \percent{12}~~        & ~~~\percent{7.5}                & ~~~\percent{4.4}  \\
Normalization factors         & ~~~\percent{8.8}      & \percent{15}~~                  & \percent{16}~~    \\
Luminosity                    & ~~~\percent{0.8}      & ~~~\percent{0.5}                & ~~~\percent{0.4}  \\
\bottomrule
\end{tabular}
\end{table}
 
Uncertainties in the renormalization and factorization scales are considered for all major background processes
by separately varying the scales \muR and \muF up and down by a factor of two. 
Additionally, PDF and \alphas uncertainties are considered by following the PDF4LHC15 prescription~\cite{Butterworth:2015oua}.  
The PDF uncertainty is evaluated as the root mean square of a set of 100 variations,
and the effect of the \alphas uncertainty is derived by taking the average difference between the up and down variations.
Additional initial-state and final-state radiation uncertainties are considered for the \ttbar and single-top processes by varying generator settings,
such as the simultaneous \muR and \muF variation and eigenvariations of the A14 tune~\cite{ATL-PHYS-PUB-2014-021}.
Furthermore, theoretical uncertainties due to the hard-scatter and parton-shower simulation are estimated
by comparing the corresponding nominal yields against those predicted with alternative generators,
\ie \POWHEG versus \AMCatNLO and \PYTHIAV{8} versus \HERWIGV{7}, respectively.
The impact of the interference between the single-top $Wt$ and \ttbar production processes is estimated
by comparing samples produced with the nominal diagram-removal scheme
with alternative samples generated with the diagram-subtraction scheme~\cite{Frixione:2008yi}.
For \Vjets, additional uncertainties related to the resummation and CKKW matching scales~\cite{Catani:2001cc,Hoeche:2009rj} are considered.
Uncertainties in the cross section and in the integrated luminosity of the data 
are applied for all simulated processes
except for \ttbar with one or two real tau leptons and single top-quark production, which are normalized to data.
The theoretical systematic uncertainties are assumed to be correlated across analysis regions
and uncorrelated between all simulated processes. 
 
\Cref{tab:SRs_unc_breakdown} summarizes the total systematic uncertainties in the background expectation in the signal regions.
In the di-tau SR the largest sources of experimental uncertainty are the uncertainties in the jet energy resolution,
whereas hard-scatter and parton-shower uncertainties dominate the uncertainty in the theoretical modeling.
For the one-bin and multi-bin SRs in the single-tau channel, 
the theoretical uncertainties in \ttbar event final-state radiation and in interference between the \ttbar and $Wt$ processes take the leading role.
 
For the theoretical uncertainties in the signal acceptance an estimate of \percent{20} is used,
which is derived from a study of the impact of varying the renormalization and factorization scales,
the radiation and merging scales,
the PDF,
and the \alphas value for several stop-stau, \LQud, and \LQv signal points.
Uncertainties in the signal production cross section are considered separately in the interpretation of the results discussed in \cref{sec:results}.
 
% End of text imported from the .//tex/systematics.tex input file
\FloatBarrier
 
\section{Results}
\label{sec:results}
% The next lines are included from the .//tex/results.tex input file

\begin{table}
\caption{
Observed event yields in data (``Observed'') and
expected event yields for SM background processes obtained from the background-only fit (``Total bkg.'' and rows below)
in the signal regions of the di-tau and single-tau channels.
The quoted uncertainties include both the statistical and systematic uncertainties and are truncated at zero yield.
By construction, no \ttTwoReal events can pass the selections in the single-tau channel.
\UncertaintiesDoNotSumUp
}
\label{tab:SR_yields}
\sisetup{parse-numbers=false}
\setlength\columnsep{0pt}
\centering
\small
\renewcommand{\arraystretch}{1.1}
\begin{tabular*}{\textwidth}{@{\extracolsep{\fill}}ll@{\hspace{-6pt}}ll@{\hspace{-6pt}}ll@{\hspace{-6pt}}ll@{\hspace{-6pt}}ll@{\hspace{-6pt}}l}
\toprule
& \multicolumn{2}{c}{Di-tau SR}         & \multicolumn{2}{c}{Single-tau SR}         & \multicolumn{6}{c}{Single-tau SR}                                                                                                  \\
&         &                             & \multicolumn{2}{c}{(one-bin)}             & \multicolumn{6}{c}{(binned in $\pt(\tau)$)}                                                                                           \\
\addlinespace[2pt]
\cline{6-11}
\addlinespace[2pt]
&         &                             &         &                                 & \multicolumn{2}{c}{$\SI{[50,100]}{\GeV}$} & \multicolumn{2}{c}{$\SI{[100,200]}{\GeV}$}  & \multicolumn{2}{c}{$>\SI{200}{\GeV}$}    \\
\midrule
Observed      & \multicolumn{2}{c}{$2$}               & \multicolumn{2}{c}{$6$}                   & \multicolumn{2}{c}{$8$}                   & \multicolumn{2}{c}{$6$}                     & \multicolumn{2}{c}{$2$}                  \\
\midrule
Total bkg.    & $4.1$   & $\pm 1.0$                   & $3.23$  & $\pm 0.55$                      & $10.1$  & $\pm 1.8$                       & $5.1$   & $\pm 1.1$                         & $2.05$    & $\pm 0.64$                   \\
\midrule
\ttTwoReal    & $0.81$  & $\pm 0.71$                  & \multicolumn{2}{c}{---}                   & \multicolumn{2}{c}{---}                   & \multicolumn{2}{c}{---}                     & \multicolumn{2}{c}{---}                  \\
\ttOneReal    & $0.82$  & $\pm 0.27$                  & $1.20$  & $\pm 0.30$                      & $4.8$   & $\pm 1.2$                       & $2.69$  & $\pm 0.88$                        & $0.64$    & $\pm 0.29$                   \\
\ttFake       & $0.51$  & $\pm 0.15$                  & $0.69$  & $\pm 0.15$                      & $2.83$  & $\pm 0.87$                      & $0.66$  & $\pm 0.17$                        & $0.185$   & $\pm 0.072$                  \\
Single top    & $0.03$  & $\substack{+0.10\\-0.03}$   & $0.39$  & $\substack{+0.45\\-0.39}$       & $0.85$  & $\substack{+0.86\\-0.85}$       & $0.54$  & $\pm 0.54$                        & $0.57$    & $\pm 0.56$                   \\
\Wjets        & $0.08$  & $\substack{+0.11\\-0.08}$   & $0.35$  & $\pm 0.16$                      & $0.34$  & $\pm 0.12$                      & $0.64$  & $\pm 0.24$                        & $0.37$    & $\pm 0.12$                   \\
\Zjets        & $0.35$  & $\pm 0.14$                  & $0.187$ & $\pm 0.054$                     & $0.275$ & $\pm 0.081$                     & $0.043$ & $\pm 0.022$                       & $0.123$   & $\pm 0.048$                  \\
Multiboson    & $0.48$  & $\pm 0.21$                  & $0.085$ & $\pm 0.037$                     & $0.163$ & $\pm 0.037$                     & $0.111$ & $\pm 0.030$                       & $0.030$   & $\substack{+0.032\\-0.030}$  \\
\ttV          & $0.60$  & $\pm 0.15$                  & $0.242$ & $\pm 0.064$                     & $0.65$  & $\pm 0.16$                      & $0.31$  & $\pm 0.12$                        & $0.092$   & $\pm 0.035$                  \\
\ttH          & $0.28$  & $\substack{+0.29\\-0.28}$   & $0.039$ & $\substack{+0.040\\-0.039}$     & $0.10$  & $\pm 0.10$                      & $0.060$ & $\substack{+0.061\\-0.060}$       & $0.028$   & $\substack{+0.029\\-0.028}$  \\
Other top     & $0.122$ & $\pm 0.067$                 & $0.043$ & $\pm 0.022$                     & $0.096$ & $\pm 0.074$                     & $0.091$ & $\pm 0.049$                       & $0.0120$  & $\pm 0.0084$                 \\
\bottomrule
\end{tabular*}
\end{table}
 
The predictions of the event yields from SM background processes obtained from the background-only fit to the control regions,
as described in \cref{sec:background},
and the observed data are shown in \cref{tab:SR_yields} for the signal regions in the di-tau and single-tau channels.
Events with pair-produced top quarks make up the largest contribution in all signal regions.
The normalization factors obtained from the background-only fit are 
$0.93\substack{+0.32 \\ -0.23}$ for the \ttTwoReal background,
$0.84\substack{+0.21 \\ -0.17}$ for \ttOneReal, and
$0.18\substack{+0.19 \\ -0.16}$ for single-top production.
The normalization factor for single-top production is significantly smaller than one
and strongly depends on how the interference between single-top production at next-to-leading order and leading-order \ttbar production
is handled~\cite{Frixione:2008yi,TOPQ-2017-05,ATL-PHYS-PUB-2016-020}.
The value $0.18$ is obtained from the samples generated with the nominal diagram-removal scheme.
The alternative diagram-subtraction scheme gives a normalization factor larger than one 
with very large uncertainties due to the much smaller yields and thus insufficient purity in the control region.
The difference between the CR yields can be attributed to the much softer \bjet distribution for the diagram-subtraction scheme.
However, the distribution shape of $\mT(\tau_1)$, the variable used in the extrapolation from the control region to the signal region in the single-tau channel,
agrees very well between the two schemes, giving confidence in the validity of the extrapolation. 
Furthermore, the predicted yields in the signal regions after the fit do not differ significantly between the two interference schemes, 
and the difference is taken into account as a systematic uncertainty.

\begin{figure}
\centering
\includegraphics[width=15cm]{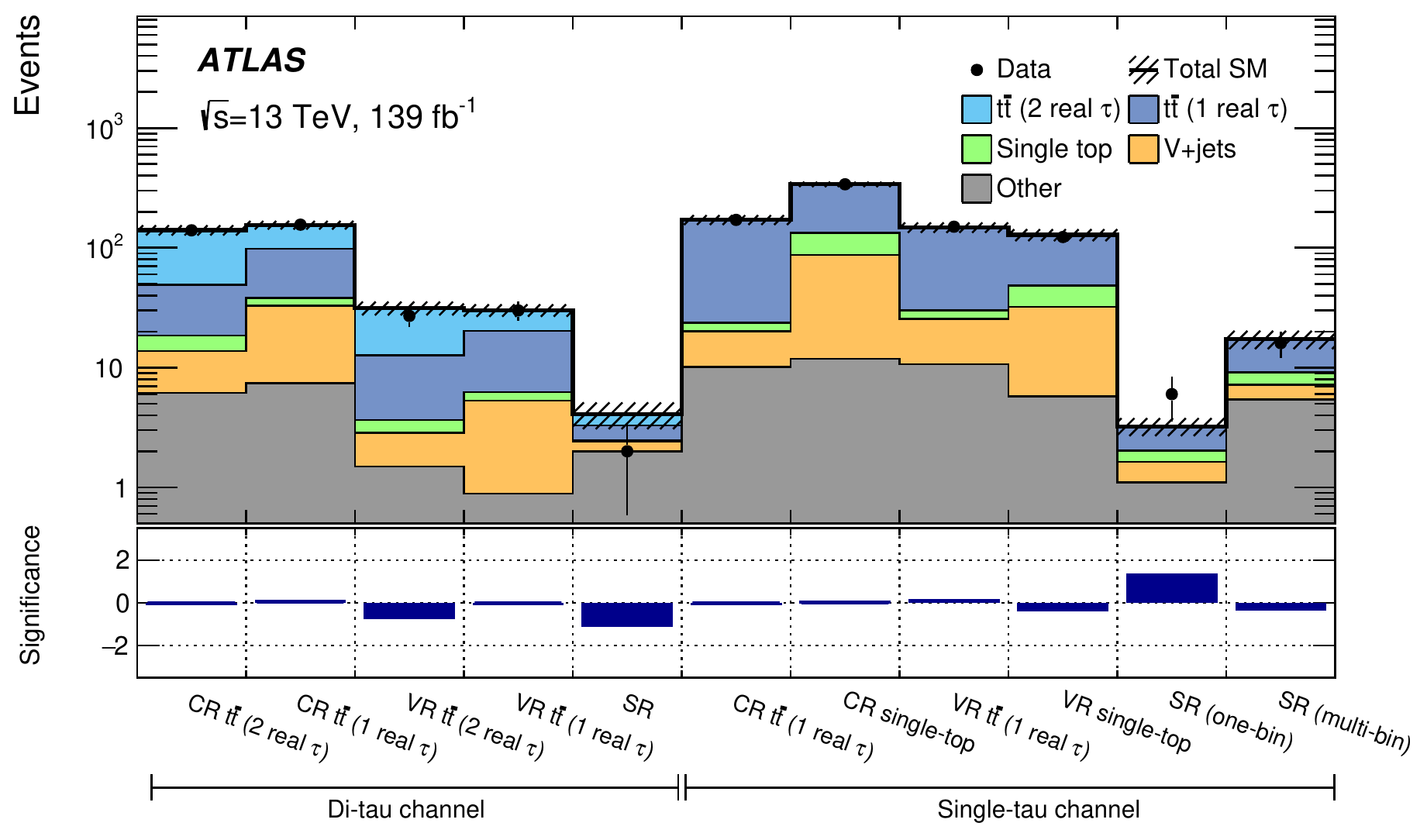}
\caption{
Comparison of expected and observed event yields (top panel)
and the significance of their difference (bottom panel)
for all analysis regions of the di-tau and single-tau channels.
The hatched band in the top panel indicates the combined statistical and systematic uncertainties in the expected SM background. 
\Postfit
\BackgroundOtherInPlots{\ttFake, \ttX, multiboson, and other top} 
The entries in the column labeled ``SR (multi-bin)'' are the sum of the three bins of the multi-bin signal region.
The significance is computed following Eq.~(25) from Ref.~\cite{Cousins:2007bmb}, 
multiplied by $-1$ if the observed number of events is smaller than the expected background yield.
}
\label{fig:bkgonlyfit_dataMC_regions_pullplot}
\end{figure}

\begin{figure}
\centering
\includegraphics[width=0.49\linewidth]{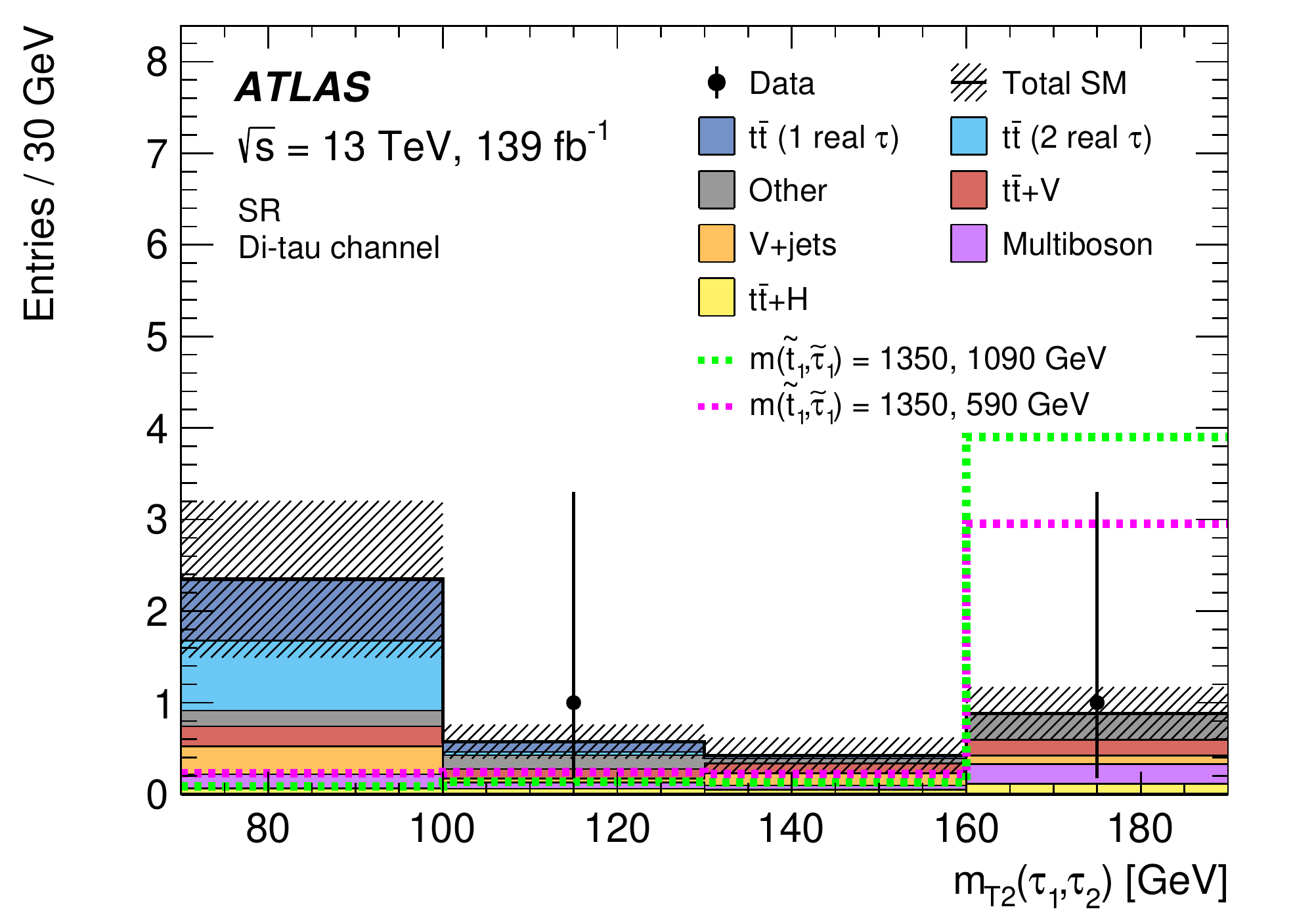}
\includegraphics[width=0.49\linewidth]{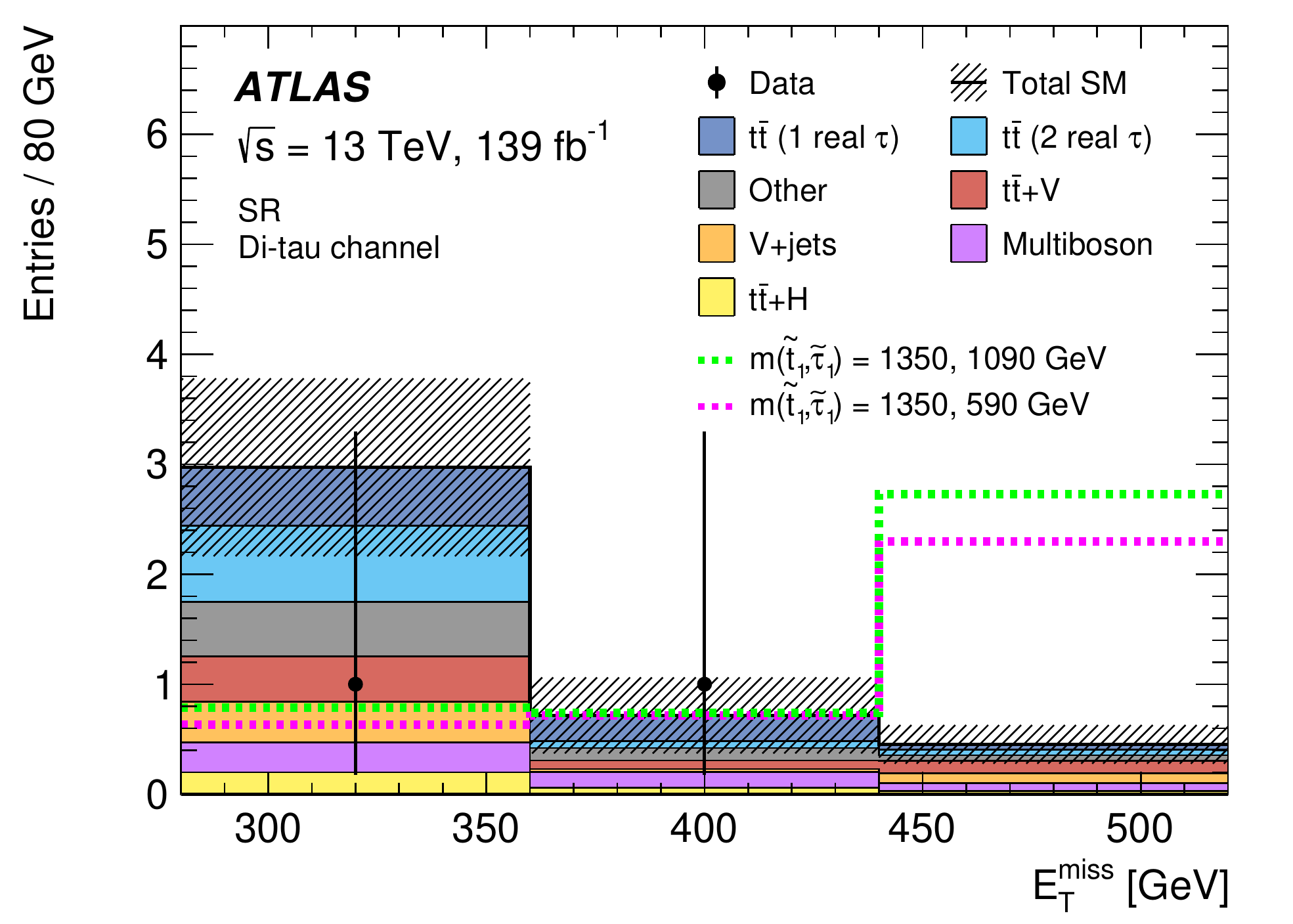}
\includegraphics[width=0.49\linewidth]{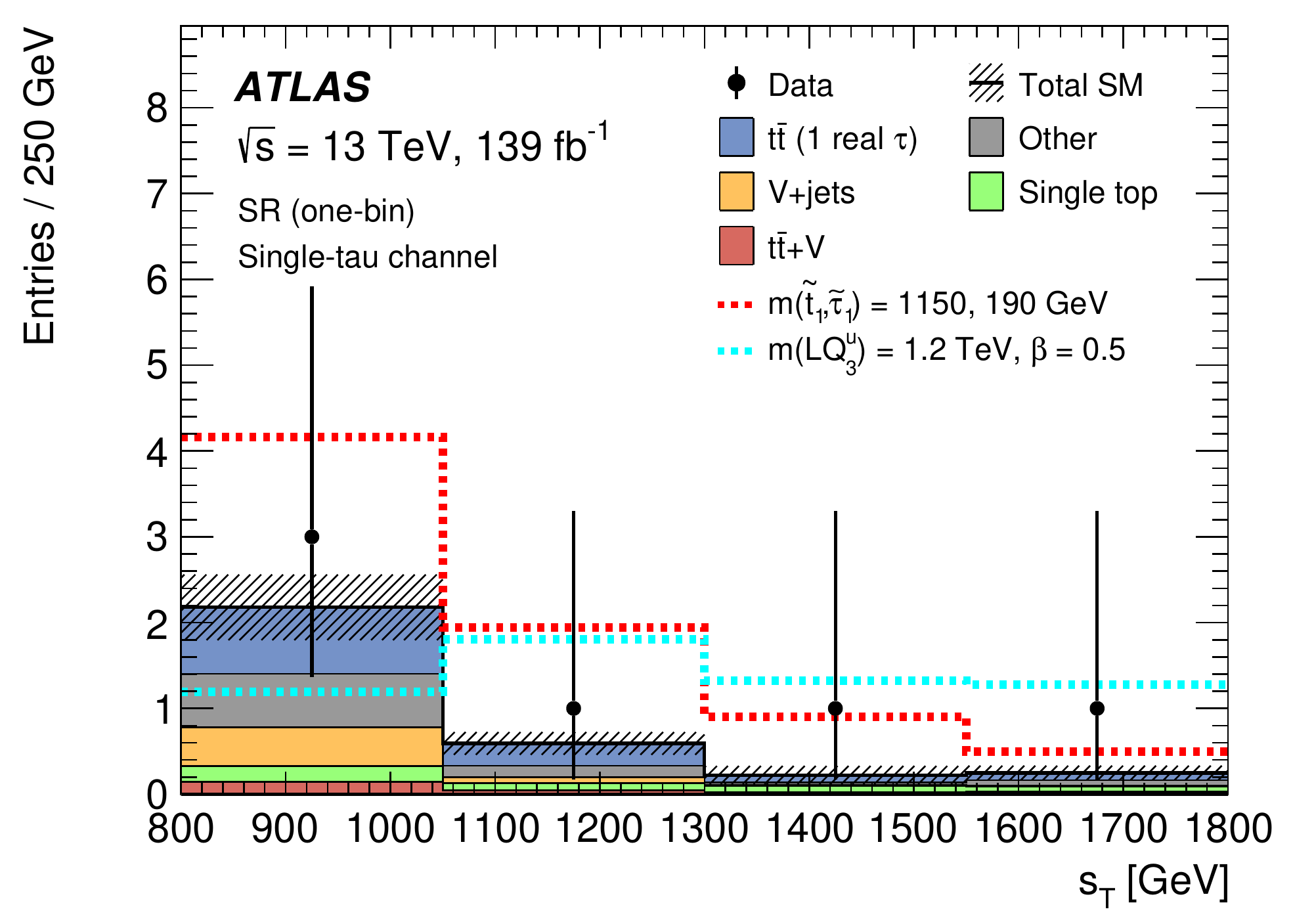}
\includegraphics[width=0.49\linewidth]{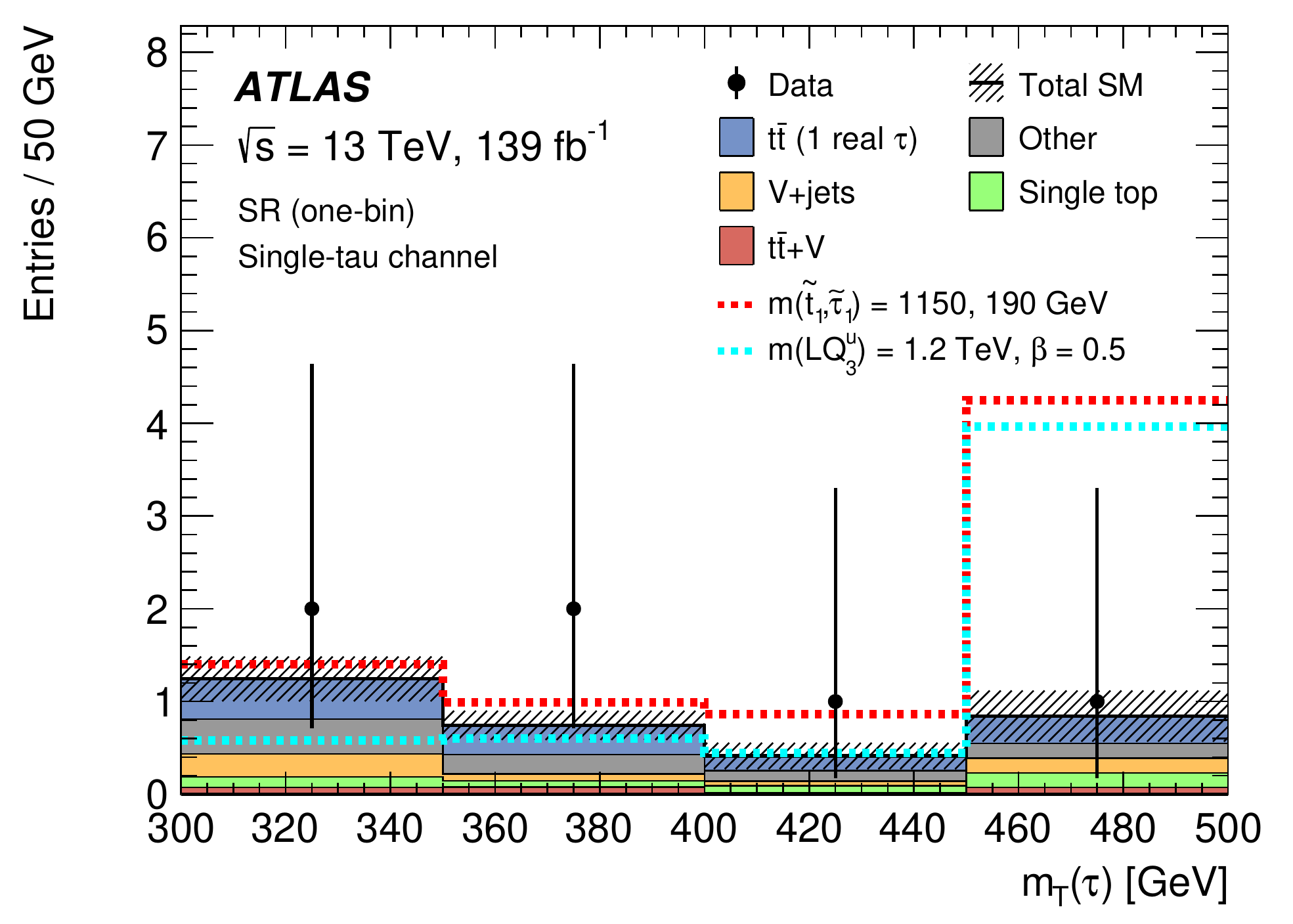}
\includegraphics[width=0.49\linewidth]{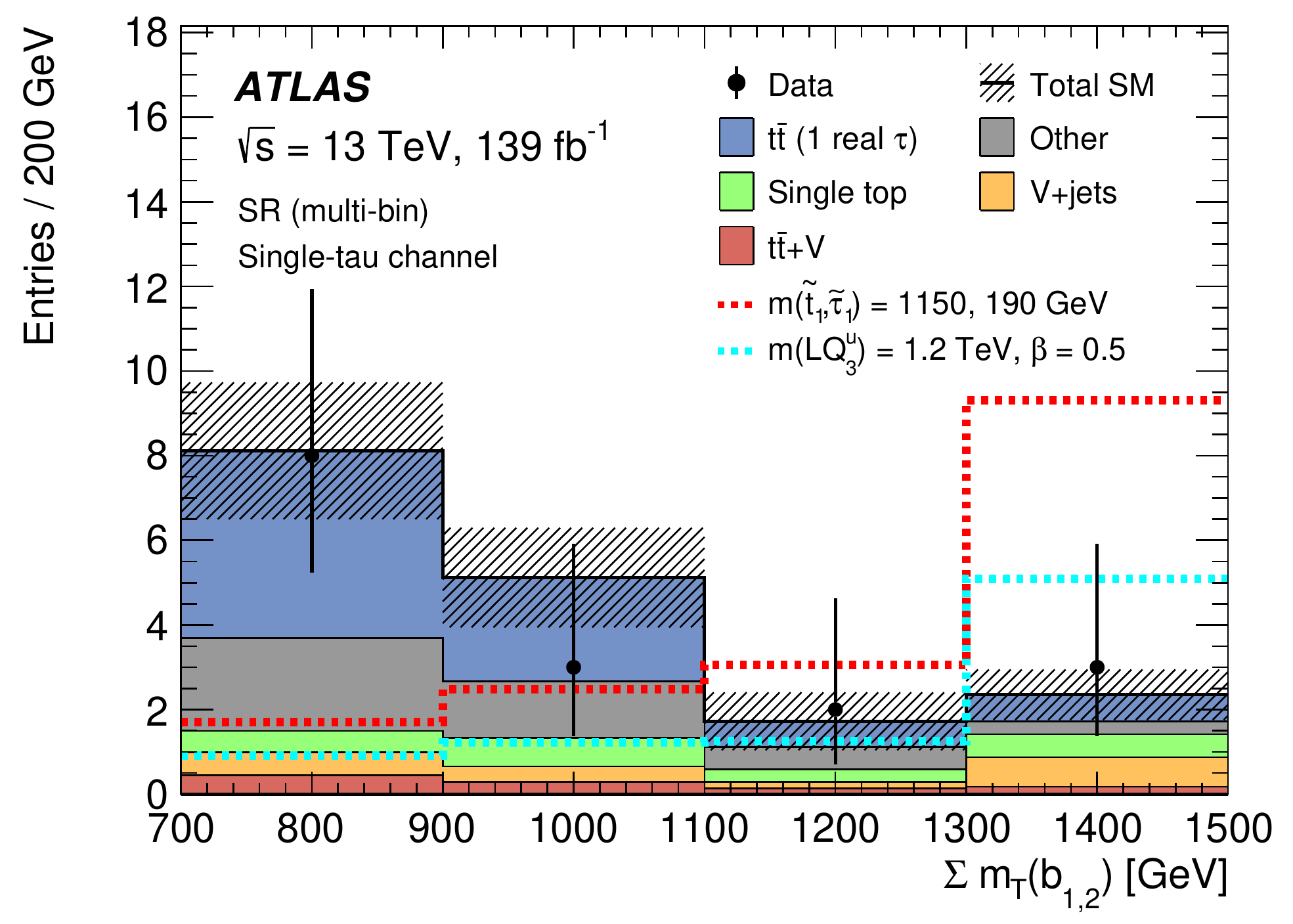}
\includegraphics[width=0.49\linewidth]{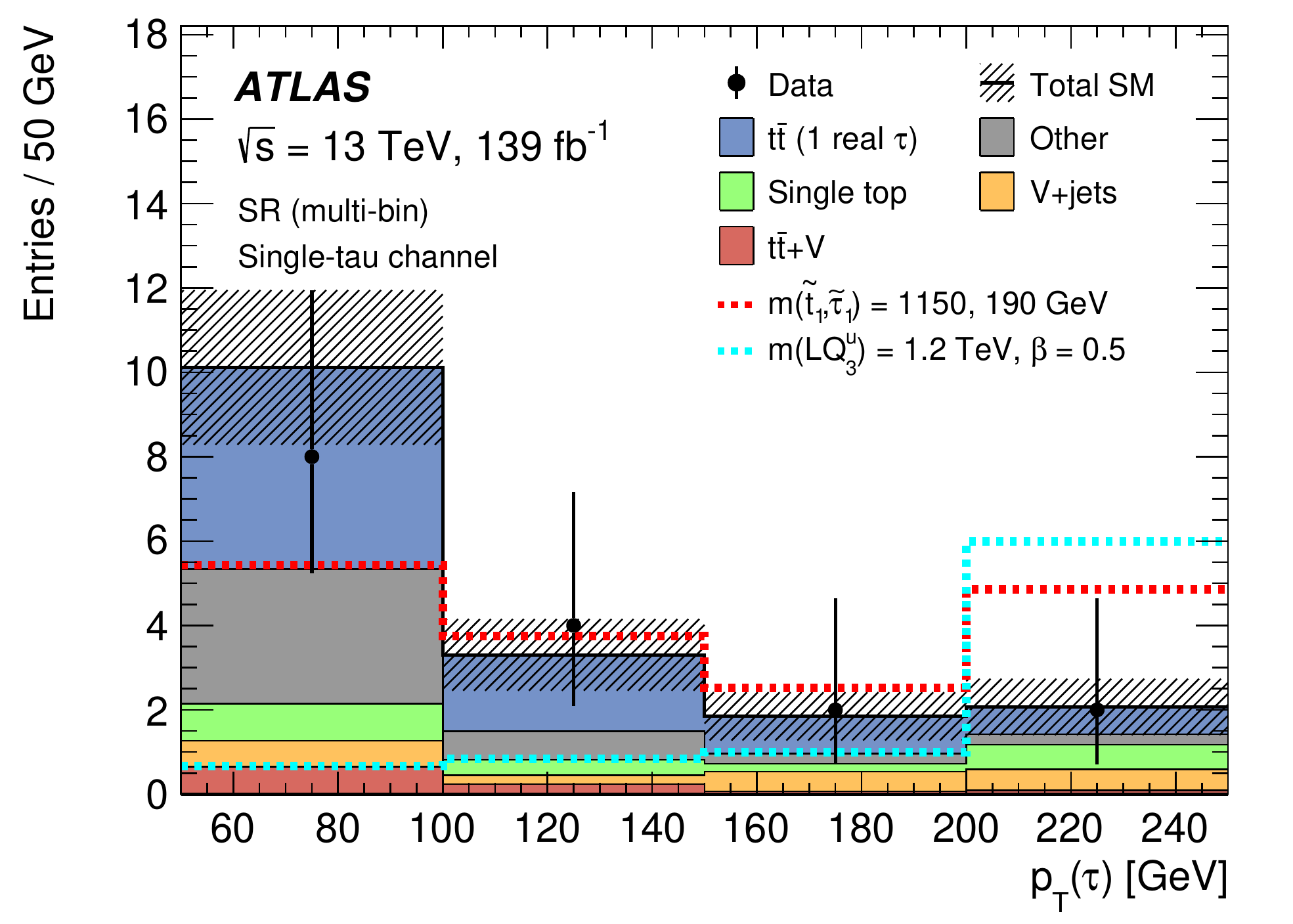}
\caption{
Distributions of $\mttwo(\tau_{1},\tau_{2})$ and \met in the di-tau SR (top),
of \st and $\mt(\tau)$ in the single-tau one-bin SR (middle)
and of \mtsumbb and $\pt(\tau)$ in the single-tau $\pt(\tau)$-binned SR (bottom).
\StackedHisto
\BandIsStatSyst
\Postfit
\BackgroundOtherInPlots{
\ttFake, single top, and other top (di-tau channel) 
or
\ttFake, \ttH, multiboson, and other top (single-tau channel)}
The overlaid dotted lines show the additional contributions for signal scenarios close to the expected exclusion contour
with the particle type and the mass and $\beta$ parameters for the simplified models indicated in the legend.
For the leptoquark signal model,
the shapes of the distributions for \LQd and \LQv (not shown) are similar to that of \LQu. 
\RightBinOverflow
}
\label{fig:SRs_dataMC}
\end{figure}

No significant excess of data events above the SM expectation is observed in any of the signal regions.
The largest excursions from the expected yields are a deficit with a significance of $1.0\sigma$ in the signal region of the di-tau channel 
and an excess with a significance of $1.3\sigma$ in the one-bin signal region of the single-tau channel,
computed with the approximate formulae from Ref.~\cite{Cowan:2010js}. 
The excess is not present, however, in the binned signal region of the single-tau channel.
\Cref{fig:bkgonlyfit_dataMC_regions_pullplot} compares the observed data yields
with the expected backgrounds for all event selections of the analysis.
The entries in the rightmost column of the plot are the sum of the three bins of the multi-bin signal region in the single-tau channel,
labeled ``SR (multi-bin)''.
\Cref{fig:SRs_dataMC} shows distributions of several kinematic variables for the expected SM background,
and compares them with the distributions expected for several benchmark signal models
and the observed data in the di-tau and single-tau signal regions.

\begin{figure}
\centering
\includegraphics[width=15cm]{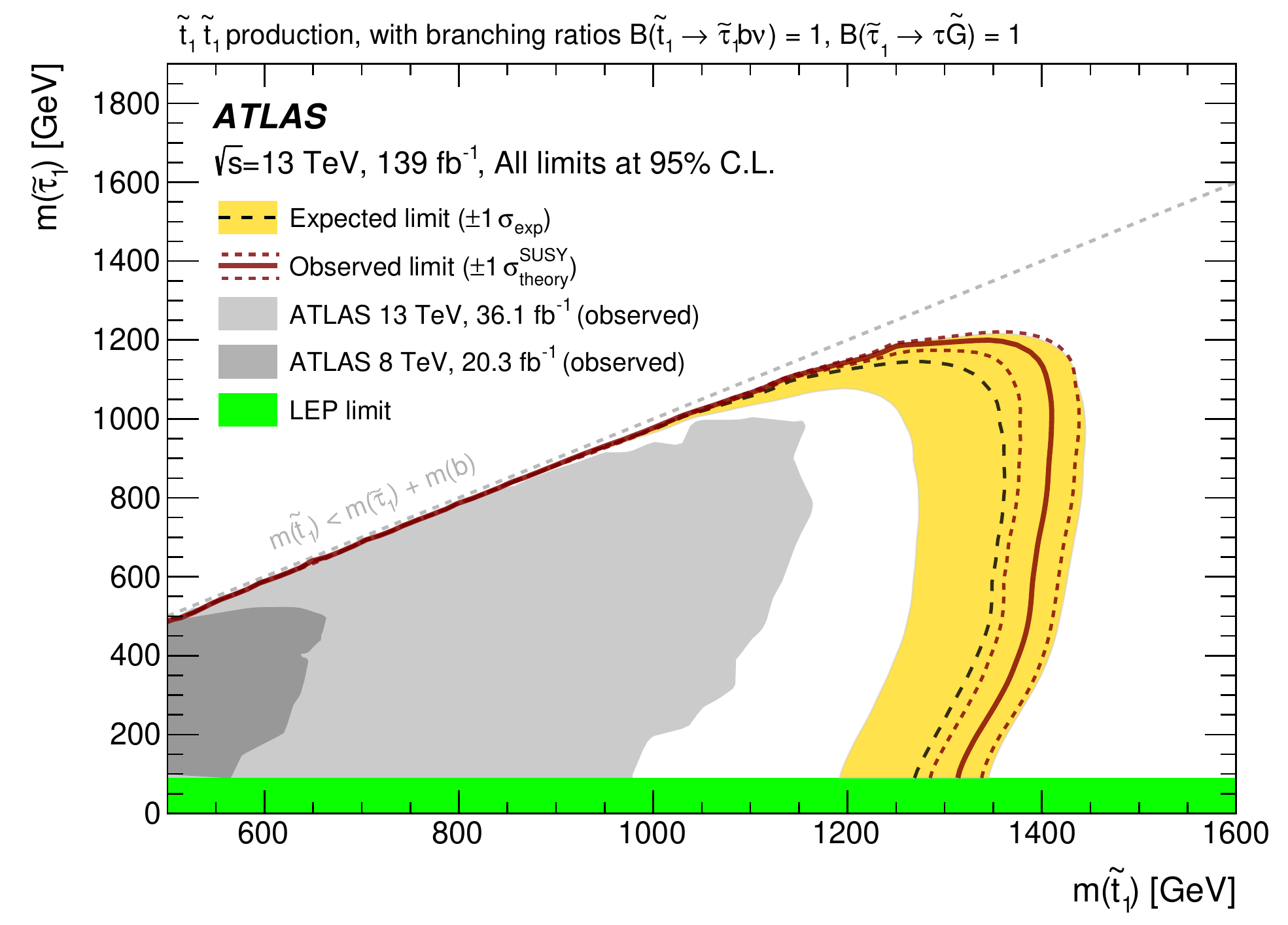}
\caption{
Exclusion contours at the \percent{95} confidence level (CL) for the stop-stau signal model
as a function of the masses of the top squark $\massof{\stop}$ and the tau slepton $\massof{\stau}$.
Expected and observed limits are shown for the present search
in comparison with observed limits from previous ATLAS analyses
based on data from Run~1 of the LHC at \eighttev~\cite{SUSY-2014-04}
and on a partial dataset from Run~2 at \tttev~\cite{SUSY-2016-19}.
The green band indicates the lower limit on the mass of the tau slepton (for a massless LSP) from the LEP experiments~\cite{web:LEP}.
}
\label{fig:contour_excl_stopstau_combined}
\end{figure}

In the absence of a significant excess,
the analysis results are interpreted in terms of exclusion limits on the parameters of the stop-stau and leptoquark signal models.
The limits are derived from a model-dependent fit,
which includes the relevant signal regions in addition to the control regions,
and signal contributions are taken into account in all analysis regions.
The signal contamination of the control and validation regions does not exceed \percent{10} (\percent{12})
for model parameters that were not excluded by previous searches for the stop-stau (scalar leptoquark) signal.
As there are no previous results for the vector-leptoquark model,
the low $\massof{\LQv}$ range is included in the interpretation,
where sizable signal contributions to the control and validation regions can be present,
exceeding \percent{10} below \GeVx{1100 (900)} for the Yang--Mills (minimal-coupling) scenario.
However, as any signal contamination of the control regions is accounted for in the model-dependent fit,
and none of the normalization factors are found to be larger than one,
the signal contamination is not expected to weaken the interpretation for these cases either.
Whether the signal-plus-background hypothesis is compatible with the observed event yields
is assessed using the \CLs prescription~\cite{Read:2002hq},
for which the $p$-values are computed with asymptotic formulae obtained for a profile-likelihood ratio as the test statistic~\cite{Cowan:2010js}.
The validity of the asymptotic formulae has been checked through a comparison with the results from pseudoexperiments
in the case of the model-independent limits.
The likelihood is the product of Poisson terms modeling the joint probability of the event yields for all analysis regions considered in the fit
and Gaussian probability terms that constrain the nuisance parameters related to the systematic uncertainties.
\Cref{fig:contour_excl_stopstau_combined} shows the expected and observed exclusion contours at the \percent{95} confidence level
for the stop-stau signal computed from the model-dependent fit that includes both the di-tau and single-tau multi-bin SR.
All systematic uncertainties are included in the fit
with the exception of the signal cross-section uncertainty,
for which a separate band around the observed limit contour is drawn instead.
The expected exclusion reach of the analysis extends to top-squark masses around \TeVx{1.35} over a wide range of tau-slepton masses,
and to tau-slepton masses around \TeVx{1.15}.
With decreasing tau-slepton mass, most noticeably below \GeVx{400}, 
the exclusion reach in top-squark mass becomes lower,
because the fraction of \met that is due to the neutrinos from the top-squark decay increases,
and thus the discrimination power of \mttwo is reduced.
The observed exclusion reach slightly exceeds the expected exclusion reach,
as the sensitivity to the stop-stau signal model for tau-slepton masses larger than \GeVx{200}
is dominated by the di-tau SR with an observed deficit.
Top-squark masses of up to \TeVx{1.4} and tau-slepton masses of up to \TeVx{1.2} are excluded at the \percent{95} confidence level in this specific model.
These are the strongest mass limits for these two supersymmetric particles in a simplified model from Run~2 of the LHC to date.
They extend significantly beyond the limits from the previous ATLAS analysis,
which are shown in the plot for comparison.
The gain in sensitivity is partly due to the larger dataset used in the analysis,
but also due to improved reconstruction and identification algorithms for tau leptons and \bjets,
and an improved signal-region strategy
with reoptimized selection requirements 
and the added single-tau signal region targeting low tau-slepton masses.

\begin{figure}
\centering
\includegraphics[width=13cm]{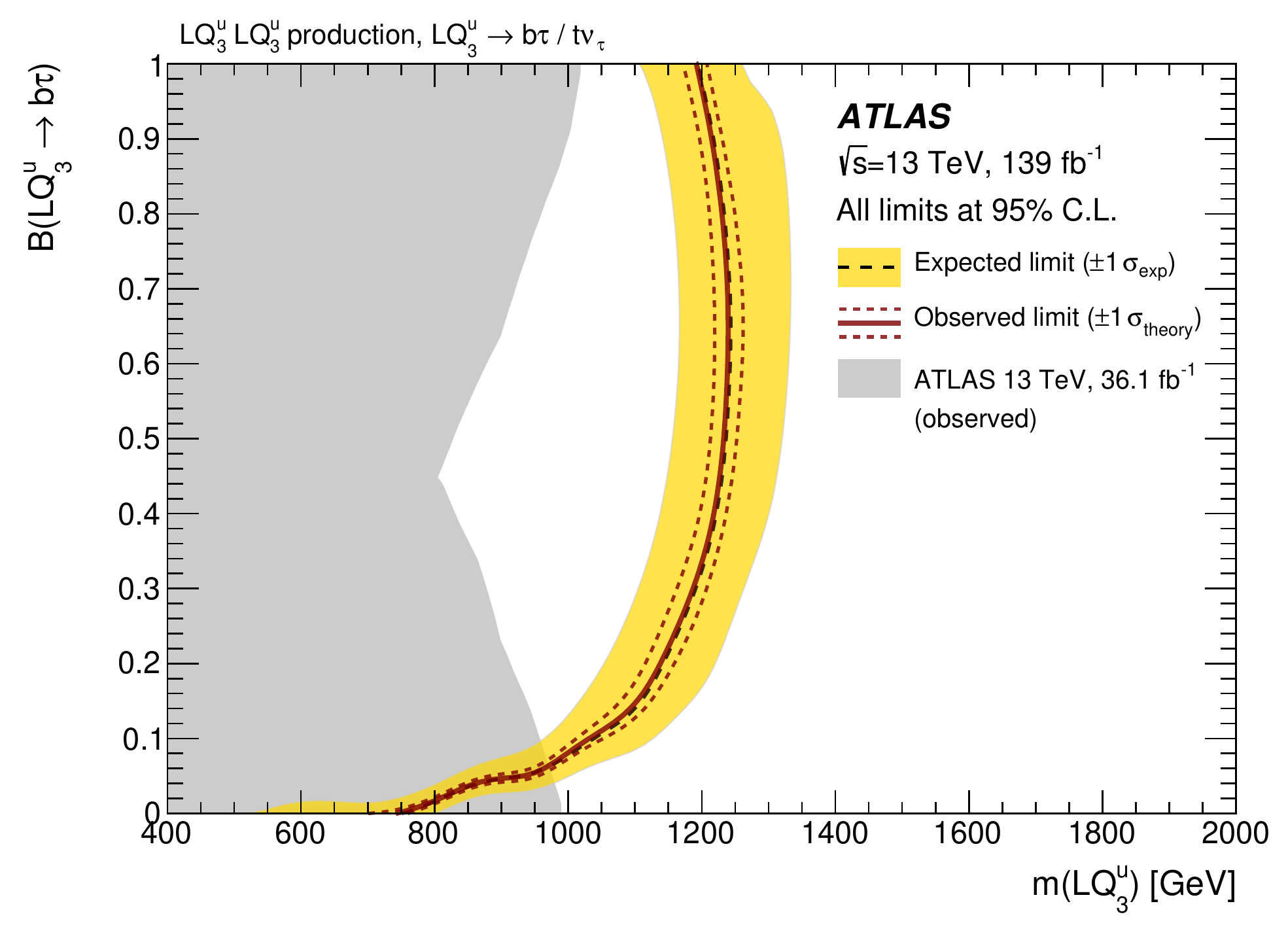}\\
\includegraphics[width=13cm]{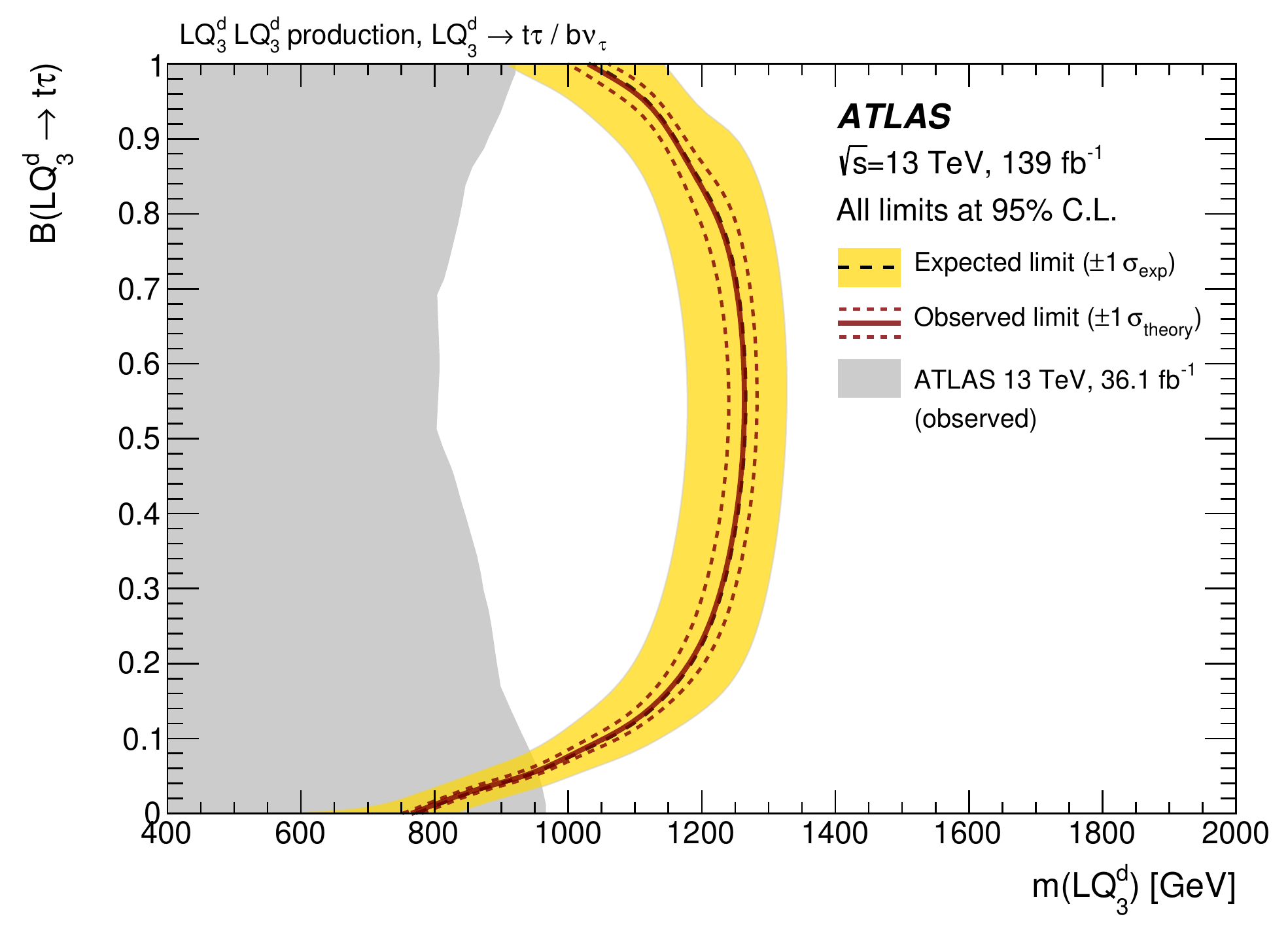}
\caption{
Expected and observed exclusion contours at the \percent{95} confidence level (CL) for the third-generation scalar-leptoquark signal model,
as a function of the mass $\massof{\LQud}$ and the branching fraction \BRcharged into a quark and a charged lepton.
The top plot shows the exclusion contour for up-type leptoquarks \LQu with charge $+\frac23e$,
the bottom plot the exclusion contour for down-type leptoquarks \LQd with charge $-\frac13e$.
The limits are derived from the binned single-tau signal region.
Shown in gray for comparison are the observed exclusion-limit contours
from the previous ATLAS publication that targets the same leptoquark models but is based on a subset of the Run-2 data~\cite{EXOT-2017-30}.
In that publication, five different analyses were considered that target not only the final state studied here
but also the final states that correspond to a branching fraction \BRcharged of 0 or 1,
leading to the concave shapes of the gray exclusion contours.
}
\label{fig:contour_excl_sLQ}
\end{figure}
 
Exclusion limits for the scalar-leptoquark signal are shown in the two plots in \Cref{fig:contour_excl_sLQ},
where the upper plot considers pair production of up-type leptoquarks \LQu
and the bottom plot pair production of down-type leptoquarks \LQd.
To derive these exclusion limits, the model-dependent fit includes, besides the four CRs, only the single-tau multi-bin SR.
The di-tau SR has not been optimized for the leptoquark models,
as final states with two tau leptons are covered by a previous search~\cite{EXOT-2019-15}. 
It has been checked that as a consequence this SR does not significantly contribute to the exclusion sensitivity, 
and it is thus not included in the interpretation for the leptoquark models.
For both types of scalar leptoquarks, the expected and observed exclusion contours extend to masses around \ourlimitsLQ at the \percent{95} confidence level
for intermediate values of the branching fraction \BRcharged.
When \BRcharged approaches zero or one,
the fraction of events with exactly one tau lepton decreases accordingly,
leading to a reduction of the signal acceptance and thus a lower mass reach. 
At $\BRcharged = 0$ the leptoquark decays do not directly produce any tau leptons.
The signal events only pass the signal-region selection
if a top quark from an up-type leptoquark decay produces a tau lepton
or if a fake tau lepton is present in the event.
This leads to a large decrease in mass reach.
The reduction of the probability of signal events to pass the SR selection when approaching $\BRcharged = 0$
is compensated by the cross section.
For $m(\LQ) = \GeVx{750}$, corresponding to the mass reach at $\BRcharged = 0$,
the cross section is larger by a factor of about 40
compared to the cross section at $m(\LQ) = \GeVx{1250}$, 
corresponding to the excluded \LQ mass at $\BRcharged = 0.5$.

\begin{figure}
\centering
\includegraphics[width=13cm]{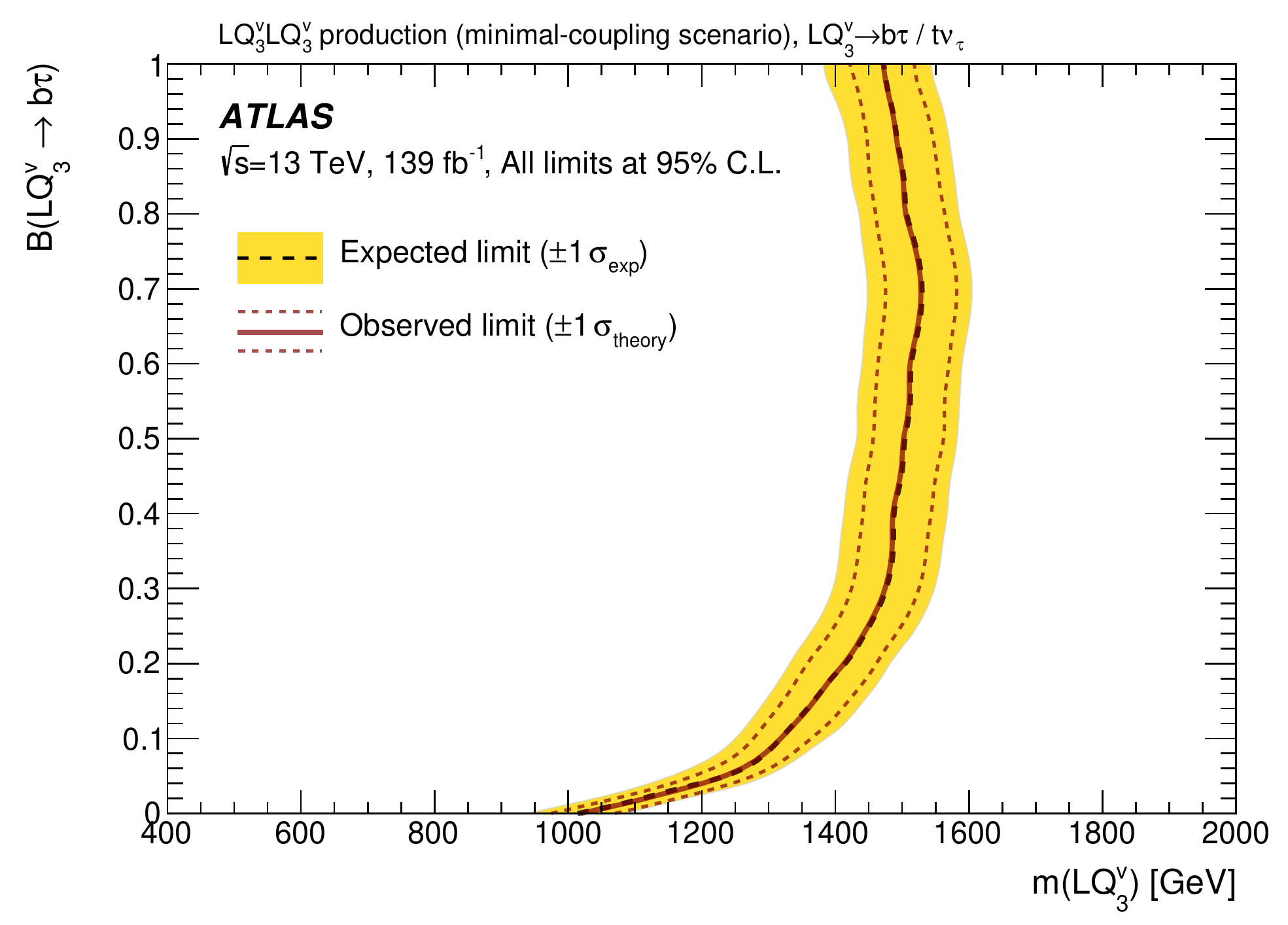}\\
\includegraphics[width=13cm]{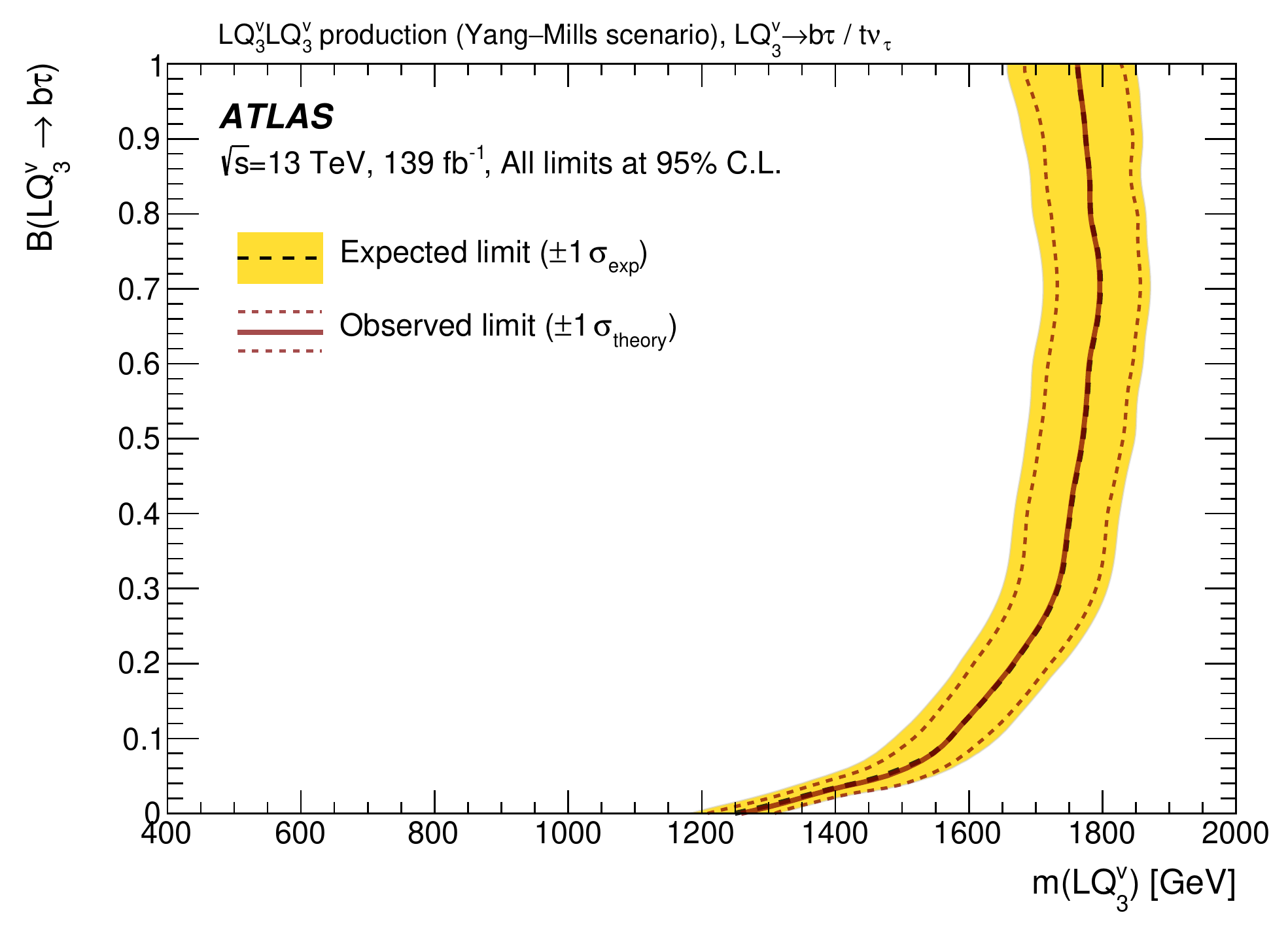}
\caption{
Expected and observed exclusion contours at the \percent{95} confidence level (CL) for the third-generation vector-leptoquark signal model,
as a function of the mass $\massof{\LQv}$ and the branching fraction \BRchargedv into a quark and a charged lepton.
The top plot shows the exclusion contour for the minimal-coupling scenario,
the bottom plot the exclusion contour for vector leptoquarks in the Yang--Mills scenario.
The limits are derived from the binned single-tau signal region.
}
\label{fig:contour_excl_vLQ}
\end{figure}
 
The exclusion contours from the interpretation of the analysis results for the vector-leptoquark models
are shown in the two plots in \Cref{fig:contour_excl_vLQ}.
As in the scalar-leptoquark case, the model-dependent fit includes, besides the four CRs, only the single-tau multi-bin SR.
For intermediate values of the branching fraction \BRchargedv,
the expected and observed exclusion contours at the \percent{95} confidence level
extend to masses around \ourlimitvLQMC in the minimal-coupling scenario
and to masses around \ourlimitvLQYM in the Yang--Mills scenario.
As expected, the shape of the contours as a function of \BRchargedv is very similar to that for the scalar-leptoquark models,
and the larger cross sections for the pair production of vector leptoquarks leads to a larger mass reach.
 
In addition to the model-dependent interpretations for the signal models shown above,
model-independent statements about the presence of physics that is not included
in the background expectation for SM processes
can also be derived from the analysis results.
The model-independent fit is performed for each of the one-bin SRs of the two analysis channels separately.
As no specific model is assumed,
the contamination of the CRs by a potential signal is neglected,
and a generic signal of variable strength is included in the SR.
\Cref{tab:modelindep_upperlimit} states
the observed and expected upper limits, $S_{\text{obs}}^{95}$ and $S_{\text{exp}}^{95}$, on the number of signal events
at the \percent{95} confidence level based on the \CLs prescription, 
where the test statistic is evaluated using pseudoexperiments.
These upper limits are also expressed as upper limits on the visible signal cross section \sigmavis,
which is defined as the product of acceptance $A$, reconstruction efficiency $\epsilon$ and signal cross section $\sigma$.
The table also reports the \CLb value, \ie the confidence level observed for the background-only hypothesis,
the discovery $p$-value,
defined as the probability to find the observed number of events or more under the background-only hypothesis,
and the equivalent significance for each of the two channels.

\begin{table}
\caption{
From left to right: upper limits at the \percent{95} confidence level (CL)
on the visible cross section ($\sigmavis$) and
on the number of signal events ($S_{\text{obs}}^{95}$).
The third column ($S_{\text{exp}}^{95}$) shows the upper limit at the \percent{95} CL on the number of signal events,
given the expected number (and $\pm 1\sigma$ excursions of the expectation) of background events.
The last two columns indicate the confidence level observed for the background-only hypothesis (\CLb),
the discovery $p$-value ($p(s=0)$)
and the significance ($Z$).
In the di-tau SR, where fewer events are observed than predicted by the fitted background estimate,
the $p$-value is capped at $0.5$.
}
\label{tab:modelindep_upperlimit}
\renewcommand{\arraystretch}{1.2}
\centering
\begin{tabular}{lcccccc}
\toprule
Analysis region         & \sigmavis~[fb]  & $S_{\text{obs}}^{95}$ & $S_{\text{exp}}^{95}$       & $\CLb$  & $p(s=0)$  & $Z$    \\
\midrule
Di-tau SR               & $0.03$          & $4.1$                   & $ { 5.3 }^{ +2.2 }_{ -1.5 }$  & $0.18$  & $ 0.50$   & $0.0$~~  \\
Single-tau one-bin SR   & $0.06$          & $8.2$                   & $ { 5.1 }^{ +2.1 }_{ -1.3 }$  & $0.91$  & $ 0.08$   & $1.37$  \\
\bottomrule
\end{tabular}
\end{table}
% End of text imported from the .//tex/results.tex input file
\FloatBarrier
 
\section{Conclusion}
\label{sec:conclusion}
% The next lines are included from the .//tex/conclusion.tex input file

In this \paperornote, a search for 
new phenomena
in final states with hadronically decaying tau leptons, \bjets and large missing transverse momentum is presented.
This signature provides sensitivity to models in which the new particles preferentially decay into third-generation Standard Model particles.
The analysis exploits the full dataset recorded with the ATLAS detector in Run 2 of the LHC,
corresponding to \SI{139}{\ifb} of proton--proton collisions at $\sqrt{s} = \SI{13}{\TeV}$.
No significant excess of events is observed over the Standard Model expectation.
The results are thus interpreted in terms of exclusion limits at \percent{95} confidence level
for two simplified models with pair production of supersymmetric top squarks or leptoquarks
which are assumed to only decay into third-generation fermions.
In the case of the supersymmetric model,
masses up to \SI{1.4}{\TeV} are excluded for top squarks
decaying via tau sleptons into nearly massless gravitinos
across a wide range of tau-slepton masses.
For both up-type and down-type scalar leptoquarks,
masses up to about \ourlimitsLQ can be excluded.
For vector leptoquarks with minimal couplings,
masses up to about \ourlimitvLQMC can be excluded,
and up to about \ourlimitvLQYM for vector leptoquarks with additional couplings to gauge bosons. 
The larger dataset,
updated reconstruction and identifications algorithms for tau leptons and \bjets,
and optimized analysis strategy
yield significantly better sensitivity than in earlier LHC studies.
Based on the considered benchmark models,
the search yields the strongest mass limits to date
on pair-produced top squarks
and on pair-produced third-generation scalar and vector leptoquarks 
at intermediate values of the branching fraction into a quark and a charged lepton.
 
% End of text imported from the .//tex/conclusion.tex input file

\section*{Acknowledgments}
% The next lines are included from the .//acknowledgements/Acknowledgements.tex input file

We thank CERN for the very successful operation of the LHC, as well as the
support staff from our institutions without whom ATLAS could not be
operated efficiently.
 
We acknowledge the support of
ANPCyT, Argentina;
YerPhI, Armenia;
ARC, Australia;
BMWFW and FWF, Austria;
ANAS, Azerbaijan;
SSTC, Belarus;
CNPq and FAPESP, Brazil;
NSERC, NRC and CFI, Canada;
CERN;
ANID, Chile;
CAS, MOST and NSFC, China;
Minciencias, Colombia;
MSMT CR, MPO CR and VSC CR, Czech Republic;
DNRF and DNSRC, Denmark;
IN2P3-CNRS and CEA-DRF/IRFU, France;
SRNSFG, Georgia;
BMBF, HGF and MPG, Germany;
GSRI, Greece;
RGC and Hong Kong SAR, China;
ISF and Benoziyo Center, Israel;
INFN, Italy;
MEXT and JSPS, Japan;
CNRST, Morocco;
NWO, Netherlands;
RCN, Norway;
MNiSW and NCN, Poland;
FCT, Portugal;
MNE/IFA, Romania;
JINR;
MES of Russia and NRC KI, Russian Federation;
MESTD, Serbia;
MSSR, Slovakia;
ARRS and MIZ\v{S}, Slovenia;
DSI/NRF, South Africa;
MICINN, Spain;
SRC and Wallenberg Foundation, Sweden;
SERI, SNSF and Cantons of Bern and Geneva, Switzerland;
MOST, Taiwan;
TAEK, Turkey;
STFC, United Kingdom;
DOE and NSF, United States of America.
In addition, individual groups and members have received support from
BCKDF, CANARIE, Compute Canada and CRC, Canada;
COST, ERC, ERDF, Horizon 2020 and Marie Sk{\l}odowska-Curie Actions, European Union;
Investissements d'Avenir Labex, Investissements d'Avenir Idex and ANR, France;
DFG and AvH Foundation, Germany;
Herakleitos, Thales and Aristeia programmes co-financed by EU-ESF and the Greek NSRF, Greece;
BSF-NSF and GIF, Israel;
Norwegian Financial Mechanism 2014-2021, Norway;
La Caixa Banking Foundation, CERCA Programme Generalitat de Catalunya and PROMETEO and GenT Programmes Generalitat Valenciana, Spain;
G\"{o}ran Gustafssons Stiftelse, Sweden;
The Royal Society and Leverhulme Trust, United Kingdom.
 
The crucial computing support from all WLCG partners is acknowledged gratefully, in particular from CERN, the ATLAS Tier-1 facilities at TRIUMF (Canada), NDGF (Denmark, Norway, Sweden), CC-IN2P3 (France), KIT/GridKA (Germany), INFN-CNAF (Italy), NL-T1 (Netherlands), PIC (Spain), ASGC (Taiwan), RAL (UK) and BNL (USA), the Tier-2 facilities worldwide and large non-WLCG resource providers. Major contributors of computing resources are listed in Ref.~\cite{ATL-SOFT-PUB-2021-003}.
 
% End of text imported from the .//acknowledgements/Acknowledgements.tex input file

\printbibliography

\clearpage % ATLAS Collaboration author list
% Reference date of SUSY-2019-18 is 2021-02-13
% Author list last updated on date 04-SEP-21
% Data extracted on 04-Sep-2021 for paper reference SUSY-2019-18
% at 8:50am
 
\begin{flushleft}
\hypersetup{urlcolor=black}
{\Large The ATLAS Collaboration}

\bigskip

\AtlasOrcid[0000-0002-6665-4934]{G.~Aad}$^\textrm{\scriptsize 99}$,    
\AtlasOrcid[0000-0002-5888-2734]{B.~Abbott}$^\textrm{\scriptsize 126}$,    
\AtlasOrcid[0000-0002-7248-3203]{D.C.~Abbott}$^\textrm{\scriptsize 100}$,    
\AtlasOrcid[0000-0002-2788-3822]{A.~Abed~Abud}$^\textrm{\scriptsize 34}$,    
\AtlasOrcid[0000-0002-1002-1652]{K.~Abeling}$^\textrm{\scriptsize 51}$,    
\AtlasOrcid[0000-0002-2987-4006]{D.K.~Abhayasinghe}$^\textrm{\scriptsize 91}$,    
\AtlasOrcid[0000-0002-8496-9294]{S.H.~Abidi}$^\textrm{\scriptsize 27}$,    
\AtlasOrcid[0000-0002-8279-9324]{O.S.~AbouZeid}$^\textrm{\scriptsize 38}$,    
\AtlasOrcid[0000-0001-5329-6640]{H.~Abramowicz}$^\textrm{\scriptsize 159}$,    
\AtlasOrcid[0000-0002-1599-2896]{H.~Abreu}$^\textrm{\scriptsize 158}$
% ...
\bigskip
\\

% ...
$^{27}$Physics Department, Brookhaven National Laboratory, Upton NY; United States of America.\\
% ...
$^{34}$CERN, Geneva; Switzerland.\\
% ...
$^{38}$Niels Bohr Institute, University of Copenhagen, Copenhagen; Denmark.\\
% ...
$^{51}$II. Physikalisches Institut, Georg-August-Universit\"{a}t G\"ottingen, G\"ottingen; Germany.\\
% ...
$^{91}$Department of Physics, Royal Holloway University of London, Egham; United Kingdom.\\
% ...
$^{99}$CPPM, Aix-Marseille Universit\'e, CNRS/IN2P3, Marseille; France.\\
$^{100}$Department of Physics, University of Massachusetts, Amherst MA; United States of America.\\
% ...
$^{126}$Homer L. Dodge Department of Physics and Astronomy, University of Oklahoma, Norman OK; United States of America.\\
% ...
$^{158}$Department of Physics, Technion, Israel Institute of Technology, Haifa; Israel.\\
$^{159}$Raymond and Beverly Sackler School of Physics and Astronomy, Tel Aviv University, Tel Aviv; Israel.\\
% ...

\end{flushleft}

% Created with Glance <Atlas.Glance@cern.ch>

\end{document}